\documentclass[11pt]{article}
\pdfoutput = 1

\usepackage[utf8]{inputenc}
\usepackage{arcs}
\usepackage{color,graphicx}
\usepackage{amsmath}
\usepackage{amssymb}
\usepackage{physics}
\usepackage{cite}
\usepackage{array}
\usepackage{setspace}
\usepackage{float}
\usepackage{tikz}
\usepackage{graphicx}
\usepackage{subfigure}
\usepackage{fancybox}
\usepackage{tikz}
\usepackage{cite}
\usepackage{mathrsfs}
\usepackage{enumitem}
\usepackage{caption}

\allowdisplaybreaks

\usepackage[margin = 2.2cm]{geometry}
\usepackage{hyperref}
\hypersetup{
    colorlinks,
    citecolor=blue,
    filecolor=black,
    linkcolor=blue,
    urlcolor=blue,
    linktocpage=true
}

\def\nn{\nonumber}

\newcommand\underrel[2]{\mathrel{\mathop{#2}\limits_{#1}}}

\newcommand{\be}{\begin{equation}}
\newcommand{\ee}{\end{equation}}
\newcommand{\bea}{\begin{align}}
\newcommand{\eea}{\end{align}}
\newcommand{\bi}{\begin{itemize}}
\newcommand{\ei}{\end{itemize}}

\newcommand{\lr}[1]{\left( #1 \right)}

\def\a{\alpha}
\def\ve{\varepsilon}
\def\nn{\nonumber}
\def\rb{\rangle}
\def\lb{\langle}
\def\ol{\overline}
\def\pd{\partial}
\def\dz{d z d \ol{z}}
\def\t{\text}
\newcommand{\lrb}[1]{\langle #1 \rangle}

\numberwithin{equation}{section}

\title{Template}

\begin{document}

\thispagestyle{empty}
\begin{center}
\vspace*{.4cm}
     {\LARGE \bf 
  Revisiting the second order formalism of JT gravity
  }
    
    \vspace{0.4in}
    {\bf Guanda Lin${}^1$, Mykhaylo Usatyuk${}^{1,2}$}
  \bigskip \rm

{\small ${}^1$ Center for Theoretical Physics and Department of Physics, Berkeley, CA, 94720, USA}\\
\vspace{0.2cm}
{\small ${}^2$ Kavli Institute for Theoretical Physics, Santa Barbara, CA 93106, USA}  \\

  \bigskip \rm
\bigskip
 
\texttt{geoff\_guanda\_lin@berkeley.edu, musatyuk@kitp.ucsb.edu}
\rm

\end{center}

\vspace{0.4in}
\begin{abstract}
We revisit the gravity path integral formalism of JT gravity. We explain how to gauge fix the path integral in the presence of asymptotic boundaries and conical defects, and resolve an ambiguity regarding the dilaton gravity operator that creates a conical defect. Along the way we study JT gravity coupled to matter on surfaces with defects of special opening angles, obtaining expressions for partition and two-point functions of matter fields. The two point function involves a summation over all geodesics on the surface, including self-intersecting geodesics, which we formally manage to include.

\end{abstract}
\pagebreak
\setcounter{page}{1}
\tableofcontents

\newpage

\section{Introduction}

Jackiw-Teitelboim (JT) gravity is a simple two dimensional model that has been central to many recent developments in quantum gravity\cite{Jackiw:1984je,Teitelboim:1983ux,Almheiri:2014cka,Maldacena:2016upp,Jensen:2016pah,Engelsoy:2016xyb,Saad:2019lba,Penington:2019kki,Almheiri:2019qdq,Iliesiu:2022kny,Mertens:2022review}. In this paper we will be revisiting various unresolved aspects of the second order formalism of the JT gravity path integral. 

The most common approach to JT gravity is through the first order formalism \cite{Saad:2019lba,Stanford:2019vob}, especially when dealing with the issue of gauge fixing the path integral. There has been some progress on understanding JT gravity from the second order formalism. For compact surfaces the path integral was properly gauge fixed in \cite{Saad:2019lba}, whereas for surfaces with asymptotically AdS boundaries significant progress was made in \cite{Moitra:2021uiv,Charles:2019tiu,Choi:2021nnq,Choi:2023syx}. Furthermore, when considering matter coupled to JT it is easiest to use gravitational variables. 

In this paper we will consider JT gravity on surfaces with asymptotically AdS boundaries with the addition of conical defects\cite{Witten:2020wvy,Maxfield:2020ale,Turiaci:2020fjj,Eberhardt:2023rzz}. We will calculate determinants of Laplace operators, matter field correlators, and gauge fix the gravity path integral on such surfaces from the perspective of the second order formalism.

Determinants of Laplace operators have recently appeared in studies of JT gravity coupled to matter where they correspond to partition functions of matter fields on the geometry\cite{Moitra:2021uiv,Jafferis:2022uhu,Jafferis:2022wez}, and in the evaluation of Lorentzian JT gravity amplitudes with topology changing wormholes \cite{Usatyuk:2022afj,Blommaert:2023vbz}. Similarly, correlation functions of matter fields on wormhole geometries have recently appeared in \cite{Saad:2019pqd,Iliesiu:2021ari,Stanford:2022fdt,Jafferis:2022wez,Jafferis:2022uhu}, where they probe the length of the wormhole. Conical defect geometries have played an important role in the counting of black hole microstates\cite{Iliesiu:2022kny}, and matter field correlators on defect geometries were considered in \cite{Iliesiu:2021ari,Boruch:2023trc}. The procedure for gauge fixing the gravitational path integral is well known from the string perturbation theory literature\cite{DHoker:1985een,Moore:1985ix,HokerPhong88}, and requires the evaluation of various determinants on hyperbolic surfaces. This naturally connects the problem of gauge fixing the path integral to understanding matter minimally coupled to JT gravity. 

We will find that the second order formalism has certain advantages. A proper gauge fixing will resolve an ambiguity regarding the correct definition of the conical defect creation operator. Furthermore, it is straightforward to incorporate matter, and we obtain closed-form expressions for determinants and two-point functions of matter coupled to JT gravity. We now summarize our main results.

\subsection*{Summary of results}

In \textbf{section \ref{sec:2}} we begin by explaining how to construct constant negative curvature surfaces by taking a quotient of the upper half-plane by a group $\Gamma \subset \text{PSL}(2,\mathbb{R})$, giving a surface $\Sigma = \mathbb{H}/\Gamma$. Using the quotient construction a variety of hyperbolic surfaces can be constructed including compact surfaces, surfaces with asymptotic AdS boundaries, and surfaces that include conical defects with special opening angles $\theta = \frac{2 \pi}{n}$ with integer $n \geq 2$. We then explain how to compute determinants of Laplace type operators on surfaces obtained through the quotient method. The final result is the following.
\begin{figure}
    \centering
    \includegraphics[width=13.5cm]{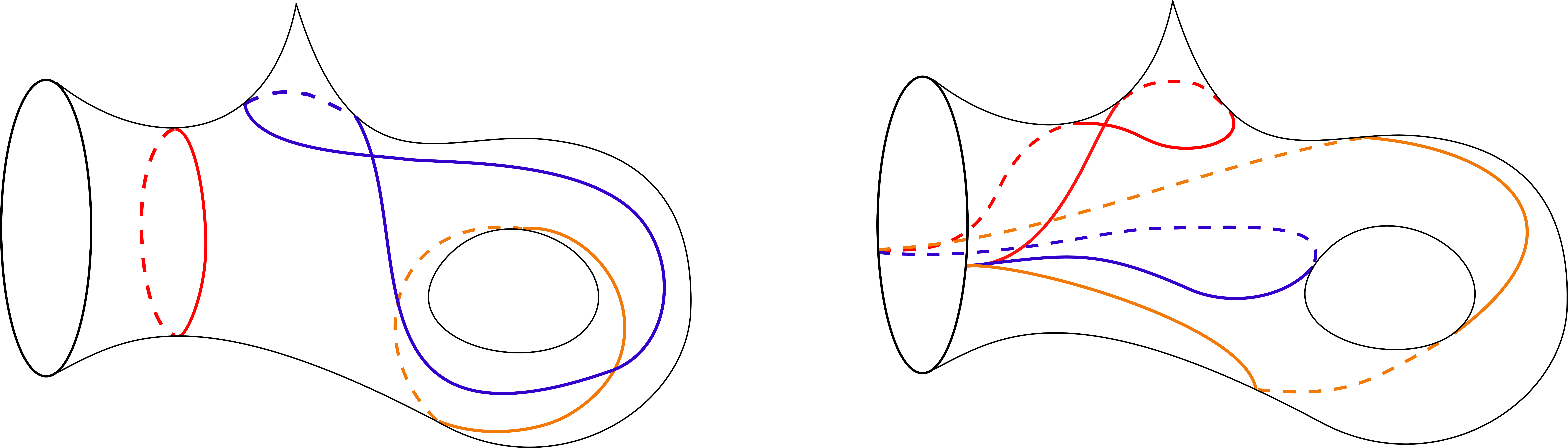}
    \caption{The determinant and two point function are reproduced by a sum over all geodesics on the surface $\Sigma$. On the left we have highlighted three geodesics that contribute to the determinant on the surface with one handle and one defect. On the right we have shown some geodesics that contribute to the two-point function, with the geodesics ending on the operator insertions at the boundary. There are infinitely many additional geodesics on these surfaces that contribute to these quantities.}
    \label{fig:introfig}
\end{figure}
\paragraph{Determinants.} Let $\Sigma=\mathbb{H}/\Gamma$ be a hyperbolic surface, either compact or with asymptotic AdS boundaries, with some number of conical defects. Define the scalar Laplacian $\Delta = - g^{a b} \nabla_a \nabla_b$. The determinant of the Laplace operator on the surface $\Sigma$ is given by
\be 
    \det\lr{{\Delta}+s(s-1)}_{\Sigma}= \underbrace{Z_{\text{hyp.}}(s)}_{\text{geodesics}} \underbrace{Z_{\text{ell.}}(s)}_{\text{defects}} \lr{\t{const}},
\ee
see section \ref{sec:2.2dets} and appendix \ref{app:determinant} for additional technical details. The most important contribution to the determinant is the Selberg zeta function\footnote{For compact surfaces this form of the determinant is well known \cite{DHoker:1986eaw,sarnak1987determinants}, and comes from the Selberg trace formula. For surfaces with asymptotically AdS boundaries the techniques for calculating determinants are relatively recent \cite{borthwick2003determinants,borthwick2005selberg,borthwick2007spectral}, and one of the new results is to incorporate conical defects in the calculation, alongside evaluating the determinant for the vector Laplacian $\det\lr{\Delta_1 + s(s-1)}_{\Sigma}$ in appendix \ref{app:determinant}.}
\be
    Z_{\text{hyp.}}(s)=\prod_{\ell_\gamma \in \mathcal{L}_\Sigma}
     \prod_{m=0}^{\infty} \lr{1-e^{-(s+m) \ell_\gamma}},
\ee
which is defined to be a sum over all closed geodesics $\mathcal{L}_\Sigma$ on the surface $\Sigma$, with lengths given by $\ell_\gamma$. See figure \ref{fig:introfig}. The integer $m$ counts the winding of the geodesics. The additional term $Z_{\t{ell.}}(s)$ comes from the presence of conical defects, and does not have an obvious geometric interpretation. We also obtain a similar expression for the determinant of the vector Laplacian $\det\lr{\Delta_1+s(s-1)}_\Sigma$, see appendix \ref{app:determinant}.

\paragraph{Correlators.} As an intermediate step in the determinant calculation we obtain a closed form expression for the two-point function of a scalar field $\lrb{\phi(x)\phi(y)}$ on an arbitrary surface $\Sigma$. When the points $x, y$ are sent to asymptotic AdS boundaries the geodesic approximation becomes exact, and the correlator reduces to a sum over all geodesics connecting $x$ and $y$. Dressing the operator insertions to the boundary Schwarzian, we find that the correlator for a field of mass $m^2 = s(s-1)$ on a surface $\Sigma=\mathbb{H}/\Gamma$ is given by
\be
G_{\Sigma}(\tau_1, \tau_2) =\sum_{\t{geodesics }\gamma} e^{-s \ell_\gamma} ~=\sum_{T \in \Gamma} \left(\frac{F'(\tau_1)(T\cdot F(\tau_2))'}{\lr{F(\tau_1)-T \cdot F(\tau_2)}^2}\right)^{s},
\ee
where the sum over all geodesics connecting the points is given by the sum over the group $\Gamma$ and the Schwarzian fluctuations are given by $F(\tau)$, see equation \eqref{2pt_final}.\footnote{In this formula we have the action $T \cdot F = \frac{a F + b}{c F + d}$. The formula also applies in the case that the operators are inserted on different asymptotic boundaries with independent Schwarzian fluctuations $F_i(\tau)$. See section \ref{sec:2.3examples} for an example.} This formula includes a summation over all geodesics connecting the two operators including self-intersecting geodesics, see figure \ref{fig:introfig}.

In \textbf{section \ref{sec:2.3examples}} we work out a number of examples for both determinants and correlators. We reproduce previous results for the defect and double trumpet geometries, and obtain new expressions for the two defect and handle disk geometries. 

\paragraph{Gravity path integral.} In \textbf{section \ref{sec:3}} we turn to the problem of gauge fixing the gravity path integral for JT gravity. Consider performing the JT gravity path integral on surfaces $\Sigma$ of genus $g$ with asymptotically AdS boundaries of regularized lengths $\vec{\beta}=\lr{\beta_1, \ldots, \beta_n}$, along with the insertion of $k$ conical defect operators $\mathcal{V}_\a$ giving rise to conical defects with opening angles $2\pi\lr{1-\a_i}$ specified by $\Vec{\a}=\lr{\a_1,\ldots, \a_k}$. We show that the path integral reduces to the Weil-Petersson measure on the moduli space $\mathcal{M}_{g,\vec{\a},\Vec{\beta}}$ of associated hyperbolic surfaces. More compactly,
\be \label{intro_PI}
Z_{g,{\Vec{\alpha}}} (\beta_1,\ldots, \beta_n) = \int \frac{\mathcal{D} g \mathcal{D} \Phi}{\t{V}(\t{Diff})} e^{-I_{\t JT}[g,\Phi]} \mathcal{V}_{\a_1} \ldots \mathcal{V}_{\a_k} = \int_{\mathcal{M}_{g,{\Vec{\alpha},\Vec{\beta}}}} d\lr{\t{Weil-Pet.}} e^{-I_{\t{bdy}}},
\ee
where on the right-hand side the boundary action is for the Schwarzian on the asymptotic AdS boundaries. In the above we have used determinants computed in section \ref{sec:2}.\footnote{We are only able to compute the necessary determinants for certain values of $\a$, but we argue that the reduction to the Weil-Petersson measure should occur for all $\a$.} We are now left with the problem of evaluating the integral on the right-hand side. It's well known from the string perturbation theory literature how to perform the integral over moduli space \cite{DHoker:1985een,HokerPhong88}, formally we have
\be \label{eqn:intro_measure}
\int_{\mathcal{M}} d (\t{Weil-Pet.}) \hspace{.1cm} e^{-I_{\t{bdy}}} = \underbrace{ \prod_n \int d m_n d \ol m_n}_{\t{coordinates on $\mathcal{M}$}} \underbrace{  \frac{\det \lrb{\mu ,\phi} \det \lrb{\ol{\mu},\ol \phi} }{\sqrt{\det \lrb{\phi, \phi} \det \lrb{\ol \phi, \ol \phi}}} }_{\t{WP measure}} \hspace{.1cm} e^{-I_{\t{bdy}}[m_n, \ol m_n]} .
\ee
The above equation can be understood as making a choice of coordinates $m_n$ for the space we are integrating over, and computing the Weil-Petersson measure in that set of coordinates. The quantities appearing in the measure are known as quadratic differentials $\phi$ and Beltrami differentials $\mu$, and we explain the technical details of the measure in section \ref{intmodulispace}. 

Computing the measure for a particular choice of coordinates is challenging for general surfaces, but in simple examples it is possible. In the case of the disk with and without a conical defect the measure can be evaluated, and in section \ref{sec:3.2examples} we perform the full integral over moduli space finding agreement with the standard Schwarzian calculation.

\paragraph{Conical defect operator.} With the above we are ready to determine the dilaton gravity operator $\mathcal{V}_\a$ that creates a conical defect when inserted into the path integral \eqref{intro_PI}. When a defect is added to the surface the real dimension of the moduli space increases by two, which corresponds to the two directions the defect can be moved on the surface. We consider a surface $\Sigma$ with a conical defect, and choose two of the coordinates $m_n$ for the moduli space to be the position $x$ of the defect. Using \eqref{eqn:intro_measure} we can immediately write down the correct operator to be
\be
\mathcal{V}_\a = \underbrace{\int_\Sigma d^2 x}_{\t{defect position}} \underbrace{ \frac{\lb \mu ,\phi_1 \rb \lb \ol \mu , \ol \phi_1 \rb}{\sqrt{\lb \phi_1, \phi_1 \rb \lb \ol \phi_1, \ol \phi_1 \rb}} }_{\t{measure for $x$ coords}} e^{-2\pi \lr{1-\a_i} \Phi(x)},
\ee
in the above we have reintroduced the exponential of the dilaton since in the measure $\eqref{eqn:intro_measure}$ we have already integrated out the dilaton.\footnote{When combined with the JT action, integrating out the dilaton creates a delta function source for the Ricci scalar at position $x$.} For additional details on the definition of this measure see section \ref{intmodulispace} around equation \eqref{eqn:conicaldefectmeasure}.  

This measure cannot be explicitly evaluated for a general surface $\Sigma$. However, in section \ref{sec:blunt_generalsurfaces} we explain that in the special limit that $\a\to1$, with opening angle $\theta = 2\pi \a$, we can evaluate the measure on an arbitrary surface $\Sigma$. We find that the conical defect operator takes the form
\be \label{eqn:Valpha_general_surfaces}
\t{General surface:} \qquad \lim_{\a \to 1} \mathcal{V}_{\a} = 2\pi (1-\alpha) \int_{\Sigma} d^2 x \sqrt{g(x)} e^{-2\pi (1-\a) \Phi(x)} + \mathcal{O}\lr{(1-\a)^2},
\ee
up to corrections in the $\lr{1-\a}$ expansion.\footnote{One interesting aspect is that the metric $g(x)$ in \eqref{eqn:Valpha_general_surfaces} is for the surface $\Sigma$ with no defect, which is important for our argument for the recursion relation for Weil-Petersson volumes found by \cite{Eberhardt:2023rzz}.} This is consistent with previous expectations for the form of the operator in the blunt defect limit $\a \to 1$ \cite{Maxfield:2020ale,Witten:2020wvy,Turiaci:2020fjj,Eberhardt:2023rzz}, and constitutes a gravity path integral argument for the form of the operator. In the case of the disk we can do better, and in section \ref{sec:blunt_disk_z0} We evaluate the full measure and find
\be
\t{Disk:}\qquad \mathcal{V}_\a = \pi (1-\a^2) \int d^2 x \sqrt{g(x)} e^{-2\pi (1-\a) \Phi(x)}.
\ee
In appendix \ref{app:diskalphaexpansion} we show how the calculation on the disk can be thought of as performing a resummation of the $\lr{1-\a}$ expansion appearing for general surfaces \eqref{eqn:Valpha_general_surfaces}. We thus conjecture that the general form of the conical defect operator on an arbitrary surface should take the form
\be \label{conjecture}
\textbf{Conjecture for general surfaces: }\qquad  \mathcal{V}_\a = \pi (1-\a^2) \int_{\Sigma} d^2 x \sqrt{g(x)} e^{-2\pi (1-\a) \Phi(x)},
\ee
after the resummation in \eqref{eqn:Valpha_general_surfaces} is performed. Our argument for the above form is that the operator should be independent of the surface it is inserted on, as is the case when these operators are defined as limits of minimal string operators \cite{Turiaci:2020fjj,Eberhardt:2023rzz}, and so the answer we find on the disk should carry over to other surfaces. 

\paragraph{Applications to dilaton gravity.} There are two immediate applications of the conical defect operator. The first is it clarifies the correct dilaton potential that corresponds to JT gravity coupled to a gas of conical defects. In \cite{Maxfield:2020ale,Witten:2020wvy,Turiaci:2020fjj,Eberhardt:2023rzz} JT gravity coupled to a gas of conical defects was defined by a summation over conical Weil-Petersson volumes with a coupling $\lambda$ weighing each defect. The bulk dilaton gravity action that corresponds to this theory is thus given by
\be
I[g,\Phi] = -\frac{1}{2} \int d^2 x \sqrt{g} \lr{\Phi R + 2 U (\Phi)},
\ee
with the dilaton potential following from equation \eqref{conjecture}
\be
U(\Phi) = \Phi + \pi \lambda (1-\a^2) e^{-2\pi\lr{1-\a}\Phi}.
\ee
At each order in the $\lambda$ expansion the path integral will localize onto singular hyperbolic surfaces and reproduce the appropriate Weil-Petersson volume with the required coupling $\lambda^k$ for surfaces with $k$ defects. 

A secondary application is that the conical defect operator can be used to give a gravity path integral argument for a recursion relation of Weil-Petersson volumes found by \cite{Eberhardt:2023rzz}. Namely, in the limit that one defect on a surface becomes blunt the volume becomes related to the volume without the defect through
\be 
\frac{d V_{g, m, n+1}\lr{\Vec{\alpha}_{n+1}, \Vec{b}_m}}{d \alpha_{n+1}} \bigg\rvert_{\alpha_{n+1}=1} = 4 \pi^2 \chi\lr{\Sigma} V_{g, m, n}\lr{\Vec{\alpha}_n, \Vec{b}_m}.
\ee
In section \ref{sec:blunt_generalsurfaces} we give a gravity path integral argument for this recursion relation using the conical defect operator. 

\section{Determinants and correlators on hyperbolic surfaces} \label{sec:2}



\subsection{Building hyperbolic geometries}
We now explain how to build constant negative curvature geometries with conical defects and asymptotic boundaries by taking quotients of the upper half-plane (UHP) by an appropriate group. For additional details see \cite{borthwick2007spectral}. The quotient construction is useful for calculating determinants and correlation functions on such surfaces since the method of images can be used. Consider the upper half-plane $\mathbb{H}$ with the standard AdS$_2$ metric
\be
ds^2 = \frac{d z d \overline{z}}{\lr{\operatorname{Im} z}^2} \,,
\ee
The group of isometries is given by PSL$(2,\mathbb{R})$, and elements of the group $T \in \text{PSL}(2,\mathbb{R})$ are either hyperbolic, elliptic, or parabolic. Hyperbolic elements satisfy $\Tr T > 2$, elliptic satisfy $\Tr T < 2$, while parabolic satisfy $\Tr T = 2$.
We will primarily be interested in hyperbolic and elliptic elements, and the most general form of these elements is given by
\be \label{eqn:HypEllElements}
\text{Hyperbolic: } T_{\ell}=\Lambda^{-1}\left(\begin{array}{ll}
e^{\ell/2} & 0 \\
0 & e^{-\ell/2}
\end{array}\right)\Lambda, \qquad \text{Elliptic: }  T_\theta=\Lambda^{-1}\left(\begin{array}{ll}
\cos (\frac{\theta}{2}) & -\sin (\frac{\theta}{2}) \\
\sin (\frac{\theta}{2}) & ~~\cos (\frac{\theta}{2})
\end{array}\right)\Lambda \, ,
\ee
where $\Lambda\in \text{PSL}(2,\mathbb{R})$. 
The action of $T$ on the upper half-plane is defined by
\be \label{eqn:Tmatrix}
T \cdot z = \frac{a z+b}{c z+d} \,.
\ee
As an example, setting $\Lambda = 1$ the hyperbolic element acts as $T_{\ell} \cdot z = e^\ell z$. Elliptic elements leave a fixed point in the interior which will become a bulk conical defect when we consider the quotient geometry.\footnote{Parabolic elements leave a fixed point at the asymptotic boundary. The fixed point corresponds to a conical singularity with deficit angle $2\pi$, and is known as a cusp. We will not consider cusps because the quotient method becomes much more complicated since the cusp lives on the asymptotic boundary. However, all of our claims should obviously generalize to surfaces that include cusps.}

One way to build a hyperbolic surface is to take a subgroup of the isometry group $\Gamma \subset \text{PSL}(2,\mathbb{R})$ and take the quotient of the upper half-plane by this subgroup $\Gamma \backslash \mathbb{H}$. That is, we identify two points as equivalent if $z \cong T \cdot z$ for any $T \in \Gamma$. To build a good hyperbolic surface we must restrict to Fuchsian groups $\Gamma \subset \text{PSL}(2,\mathbb{R})$, which are discrete subgroups of $\text{PSL}(2,\mathbb{R})$. This restricts the elliptic elements we are allowed to consider to $T_\theta^n = \pm \operatorname{I}$, which only allows opening angles $\theta = 2\pi /n$ with integer $n\geq2$.\footnote{Restricting to Fuchsian groups is equivalent to imposing the condition that a sufficiently small ball  around the identity element, in the $SL(2,\mathbb{R})$ group manifold, contains no other elements. Intuitively, when quotienting $z \cong T \cdot z$ we do not want there to exist elements $T\in \Gamma$ arbitrarily close to the identity, since then we will be identifying arbitrarily close points $z, T \cdot z$, and the quotient surface will be degenerate.} We will take the Fuchsian group to have a finite number of generators $ T_i $ given by some set set of hyperbolic and elliptic elements \eqref{eqn:HypEllElements}. 

We are interested in hyperbolic surfaces with both asymptotic boundaries and conical defects. Locally, a conical defect at a point $x_i$ is characterized by the fact that we can travel around the point by going through an angle $\theta< 2\pi$. For a conical defect of opening angle $\theta = 2\pi \a$ this translates to a condition on the scalar curvature given by
\be
\frac{1}{2}\sqrt{g}\lr{R+2}=2\pi \lr{1-\a}\delta^2(x-x_i)\,,
\ee
where we restrict to $\alpha \in (0,1)$. A natural question is whether there exists a hyperbolic surface with $k$ defects with angles specified by $\vec{\a}=\lr{\alpha_1,\ldots,\a_k}$. It turns out that such a surface always exists provided that the specification of the defects $\alpha_i$ does not violate the Gauss-Bonnet theorem \cite{troyanov1991prescribing,mcowen1988point}
\be
2\pi \chi(\Sigma) = \frac{1}{2}\int_\Sigma \sqrt{g} R + \int_{\partial \Sigma} \sqrt{h} K\,,
\ee
where $K$ is the extrinsic curvature on the boundary of the surface, $h$ is the induced boundary metric, and $\chi(\Sigma)=2-2g-n$ where $n$ is the number of boundaries and $g$ is the genus. Conical defects fall into two classes: sharp defects with opening angle $\theta \leq \pi$, and blunt defects with opening angle $\theta >\pi$ \cite{Turiaci:2020fjj,Eberhardt:2023rzz}. This translates to the condition $\alpha \leq \frac{1}{2}$ for sharp and $\alpha > \frac{1}{2}$ for blunt.\footnote{One nice property of sharp defects is that for surfaces with an asymptotic boundary and at least two sharp defects, the defects are separated from the boundary by a closed geodesic homotopic to the boundary. This can be seen from the Gauss-Bonnet theorem.} The quotient construction only allows us to build surfaces with sharp defects $\a \leq  \frac{1}{2}$.

We now summarize existence theorems on the types of surfaces that can be built using the quotient method. First, consider the moduli space $\mathcal{M}_g$ of compact surfaces of genus $g\geq2$. For each compact hyperbolic surface $\Sigma$ there exists a Fuchsian group $\Gamma$ such that $\Sigma = \Gamma \backslash \mathbb{H}$, and such a representation of the surface is known as a Fuchsian model. As an example, for $g=2$ the Fuchsian group is generated by $4$ hyperbolic elements $T_i$ with a non-trivial constraint\footnote{The constraint enforces that the closed loop generated by the action of the following group element will be contractible on the surface.} on the generators
\be
\Gamma = \langle T_1, T_2, T_3, T_4 ~| ~T_4^{-1}T_3^{-1} T_4 T_3 T_2^{-1}T_1^{-1} T_2 T_1 = 1 \rangle.
\ee
In general, the Fuchsian group of a compact genus $g$ surface is generated by $2 g$ distinct hyperbolic elements satisfying non-trivial constraints.

Now consider the moduli space $\mathcal{M}_{g,\Vec{\a}}$ of compact surfaces of genus $g$ with $k$ conical defects with deficit angles specified by $\vec{\a}=\lr{\a_1,\ldots, \a_k}$. When the opening angles take special values $\theta_i = 2\pi \a_i$ with $\a_i = \frac{1}{n_i}$ and integer $n_i \geq 2$ then every surface can be obtained by a quotient with a suitable Fuchsian group\cite{scott1983geometries,thurston2022geometry}. The group will contain $k$ elliptic elements that leave a fixed point $z_i \in \mathbb{H}$ which becomes the location of the conical defect.\footnote{The order of the elliptic element determines the strength of the opening angle to be $2\pi/n$.} Surfaces with other deficit angles cannot be constructed by the quotient method. 

In JT gravity we typically consider non-compact surfaces with $n$ asymptotic boundaries with regularized lengths $\vec{\beta} = \lr{\beta_1,\ldots,\beta_n}$. All hyperbolic surfaces with asymptotic boundaries, but without conical defects, $\mathcal{M}_{g, \vec \beta}$ can be obtained by a quotient with a Schottky group (see theorem 15.3 in \cite{borthwick2007spectral}). We are not aware of a similar theorem in the case that we include conical defects $\mathcal{M}_{g,\vec \a, \vec \beta}$, but we will assume that if we limit the opening angles $\Vec{\a}$ to the previously mentioned special values that these surfaces can also be obtained through a quotient.

\subsection{Two-point functions and determinants on quotient surfaces} \label{sec:2.2dets}

The quotient construction makes it possible to calculate the determinant and two-point function on the associated surface. In this subsection we explain how to evaluate these quantities.

Let $\Sigma$ be a hyperbolic surface with metric $g$ obtained by quotienting $\mathbb{H}$ with a Fuchsian group $\Gamma$. The scalar Laplacian is defined to be $\Delta = -g^{a b}\nabla_a \nabla_b$. The two-point function on the surface $\Sigma$, also known as the resolvent, is defined through the equation 
\be \label{eqn:resolvent}
\lr{\Delta +s(s-1)}R_\Sigma(s;z,w)=\frac{\delta^2(z-w)}{\sqrt{g}}\,,
\ee
where $z,w$ are two points on the surface $\Sigma$, and $s\geq 1$ is related to the mass of a scalar field through $m^2=s(s-1)$. For surfaces obtained through the quotient method we can obtain a formula for the resolvent by using the method of images. The key idea is that the resolvent on the upper half plane $R_\mathbb{H}$, which is explicitly known, locally satisfies the desired equation \eqref{eqn:resolvent}, but globally we must have that $R_\Sigma(s;z,w) = R_\Sigma(s;T \cdot z,w)$ since $z \cong T \cdot z$ for all $T\in \Gamma$. Performing a sum over the quotient group gives a function that precisely satisfies all the properties that define the resolvent on $\Sigma$ 
\be \label{eqn:imageresolvent}
R_\Sigma(s;z,w) = \sum_{T \in \Gamma} R_{\mathbb{H}}(s;T\cdot z,w).
\ee
Once the resolvent is known it is straightforward to calculate the determinant. For smooth compact surfaces it is well known that the determinant of the Laplacian reduces to a product over geodesics due to the Selberg trace formula\cite{DHoker:1986eaw,sarnak1987determinants}. However, for non-compact surfaces this method is not applicable, and other techniques need to be used\cite{borthwick2003determinants,borthwick2005selberg,borthwick2007spectral}. We will review the technique for surfaces with asymptotic boundaries\cite{borthwick2007spectral}, and extend it to include conical defects. The basic procedure of the calculation and key results will be presented here, leaving technical details to appendix \ref{app:determinant}.

To compute the determinant of an operator we must solve the eigenvalue problem $\lr{\Delta + s(s-1)}\phi_n = \lambda_n \phi_n$ and take the product of all the eigenvalues. However, doing this directly is quite challenging. A simpler technique is to define the determinant through the trace of the resolvent. As an example, consider the case that the eigenvalues are discrete with eigenfunctions $\phi_n$. The resolvent is given by
\be 
R_\Sigma(s;z,w) = \sum_n \frac{\phi_n(z) \phi_n(w)}{\lambda_n + s(s-1)}\,,
\ee
where the eigenfunctions satisfy $\int_\Sigma d^2z \sqrt{g} \phi_n(z) \phi_m(z) = \delta_{n m}$. One can check that the determinant defined as a product of eigenvalues is related to the resolvent trace through
\be \label{eqn:determinant_from_res}
    \frac{1}{2s-1}\frac{d}{ds} \log(\det(\Delta+s(s-1)))= {\rm tr}R_{\Sigma}(s)\equiv \int_{\Sigma} d^2 z \sqrt{g} R_\Sigma(s;z,z)\,.
\ee
Conversely, if a closed form expression for the resolvent can be obtained then the determinant can be obtained by inverting the above formula.

For simple surfaces the resolvent trace is easy to evaluate since the resolvent is known on the UHP, and we can perform the sum over the Fuchsian group defining the surface. Consider the double trumpet with geodesic throat of size $\ell$. The Fuchsian group is generated by a single hyperbolic element $\Gamma= \lrb{T_{\ell}}$, producing the surface $\Sigma = \mathbb{H}/\lrb{T_{\ell}}$. The resolvent trace is obtained after using equations \eqref{eqn:imageresolvent} and \eqref{eqn:determinant_from_res}
\begin{align}\label{eqn:DTresolventtrace}
\sum_{\substack{m\in \mathbb{Z}_{\ne 0}}}\int_{\mathbb{H}/\lrb{T_\ell}} d^2 z \sqrt{g} R_{\mathbb{H}}\lr{s ; T_{\ell}^m \cdot z,  z}  
=\frac{2\ell}{2s-1}\sum_{m=1}^{\infty} \frac{e^{-sm\ell}}{1-e^{-m\ell}}\,.
\end{align}
The integer $m$ counts the number of windings of the closed geodesic. 
The integral can be implemented within a fundamental domain given by $1<|z|<e^{\ell}$; as for details of integration, see appendix~\ref{app:determinant}.

For general surfaces the summation over the group \eqref{eqn:imageresolvent} will reduce the resolvent trace to simpler building blocks, such as the above calculation on the double trumpet. Let us explain how this works. The sum over the group can be broken up into a sum over conjugacy classes $[\gamma]$ of primitive elements as $\sum_{\Gamma} = \sum_{[\gamma]} \sum_{g \in \Gamma / C_{\lrb{\gamma}}}$.\footnote{A primitive element $\gamma$ is not a power of any other element in the group. The centralizer is defined as the set of elements that leave the group generated by $\gamma$ to be invariant $C_{\lrb{\gamma}} = \{ g \lrb{\gamma} g^{-1} = \lrb{\gamma} | g \in \Gamma \}$. The set of left cosets $\Gamma / C_{\lrb{\gamma}}$ generates a list of distinct elements that we should conjugate $\gamma$ by sum over the full conjugacy class. We must take the centralizer of the group generated by $\gamma$, $C_{\lrb{\gamma}}$ because we want to break up the sum into conjugacy classes of $[\gamma^n]$. It turns out that the centralizer for primitive $\gamma$ is simply given by $C_{\lrb{\gamma}}=\lrb{\gamma}$, which we use notationally throughout.} To avoid over-counting, we must sum over an element from each coset $g \in \Gamma / C_{\lrb{\gamma}}$ and conjugate $g \gamma g^{-1}$ to sum over the entire conjugacy class. The primitive elements will either be hyperbolic or elliptic which geometrically corresponds to either closed geodesics or conical defects respectively, and we must also include a sum over conjugacy classes of multiples of the primitive elements $\gamma^n$. The final result is that every group element can be reached by a conjugation $g^{-1} \gamma^n g$ of some power of a primitive element $\gamma$. Suppose we have a list of primitive elements denoted by $\Pi$, the summation over the group is then given by\footnote{When $\gamma$ is elliptic the summation over powers $\gamma^m$ should be cutoff when we reach $\gamma^m = I$.} 
\be \label{eqn:resolvent_decomposition}
   R_\Sigma(s;z,w)=R_{\mathbb{H}}\lr{s;z,w}+ \sum_{\gamma \in \Pi} \sum_{g \in \Gamma /\lrb{\gamma}} \sum_{m \neq 0}^\infty  R_{\mathbb{H}}\lr{s ;g^{-1} \gamma^m g \cdot z, w}\,,
\ee
where the first term is the identity contribution.

When calculating the trace of the resolvent \eqref{eqn:determinant_from_res} the sum over $\Gamma / \lrb{ T_{\ell}}$ for hyperbolic elements $T_{\ell}$ will transform the domain of integration from the surface $\Sigma$ to the fundamental domain of a double trumpet with a closed geodesic $\ell$, of which we already have the answer in \eqref{eqn:DTresolventtrace}
\begin{align}
\sum_{\substack{m\in \mathbb{Z}_{\ne 0}}} \sum_{g \in \Gamma / \langle T_{\ell}\rangle } \int_{\mathbb{H}/\Gamma} d^2 z \sqrt{g} R_{\mathbb{H}}\lr{s ; z, T^m_{\ell} g \cdot z, g \cdot z} &= \sum_{\substack{m\in \mathbb{Z}_{\ne 0}}} \int_{\mathbb{H}/\lrb{T_\ell}} d^2 z \sqrt{g} R_{\mathbb{H}}\lr{s ; z, T_{\ell}^m\cdot z} \nn\\
&=\frac{2\ell}{2s-1}\sum_{m=1}^{\infty} \frac{e^{-sm\ell}}{1-e^{-m\ell}}\,,
\end{align}
where the RHS is already given in \eqref{eqn:DTresolventtrace}. This gives a contribution to the determinant \eqref{eqn:determinant_from_res} for each closed geodesic $\ell$ on the surface $\Sigma$. The same logic applies to elliptic elements $T_{\theta}$ in the sum \eqref{eqn:resolvent_decomposition} except now the fundamental domain becomes a cone.


The full determinant calculation is quite involved and carried out in appendix \ref{app:determinant}. We quote the final result. Consider a hyperbolic surface $\Sigma$ with $k$ conical defects with opening angles $\theta_i = \frac{2 \pi}{n_i}$ with integer $n_i \geq 2$. The surface can be compact or with asymptotic AdS boundaries. The determinant is given by
\be \label{eqn:detfinal}
    \det\lr{{\Delta}+s(s-1)}= \underbrace{Z_{\text{hyp.}}(s)}_{\text{geodesics}} \underbrace{Z_{\text{ell.}}(s)}_{\text{defects}} \lr{\t{const}}
\ee

The most important contribution is given by $Z_{\text{hyp.}}$, which is the Selberg zeta-function on $\Sigma$. It comes from the summation over hyperbolic elements in \eqref{eqn:resolvent_decomposition}. It is defined by a product over contributions from all closed geodesics  on the surface. Define the set of lengths $\ell_\gamma$ of primitive closed geodesics to be $\mathcal{L}_\Sigma$,\footnote{Excluding exceptional cases, geodesics are endowed with an orientation. For each geodesic $\gamma$ there is a mirror geodesic with opposite orientation and the same length. Both a geodesic and it's mirror need to be independently included in the Selberg zeta-function. An example of a non-orientable geodesic is given by the two defect surface considered below.} and the Selberg zeta-function for the surface $\Sigma$ is defined to be
\be \label{eqn:selbergzeta}
    Z_{\text{hyp.}}(s)=\prod_{\ell_\gamma \in \mathcal{L}_\Sigma}
     \prod_{m=0}^{\infty} \lr{1-e^{-(s+m) \ell_\gamma}},
\ee 
which can be derived from \eqref{eqn:DTresolventtrace}.
In the above equation, the first product is over all primitive geodesics, while the second is over all integer $m$ windings of the geodesic. Primitive geodesics include self-intersecting ones as well as those that touch conical defects.
We also mention that geodesics with opposite orientations are counted as distinct in \eqref{eqn:selbergzeta}.

The second term comes from the summation over elliptic elements in \eqref{eqn:resolvent_decomposition}. Since elliptic elements give conical defects we label this the ``defect'' contribution, and it is given by
\be
    Z_{\text{ell.}}(s)=\prod_{i=1}^{k} \prod_{r=0}^{n_i-1} \Gamma\left(\frac{s+r}{n_i}\right)^{\frac{2 r+1-n_i}{n_i}}.
\ee
For each conical defect we get a highly non-analytic contribution in the opening angles $\theta_i=\frac{2\pi}{n_i}$. Unlike the Selberg zeta-function, the contribution due to elliptic elements does not seem to have an obvious geometric interpretation. We emphasize that every term in equation \eqref{eqn:detfinal} depends on the choice of deficit angles in some way. Either implicitly through the types of geodesics that are included in the Selberg zeta-function, or explicitly as some function of the angles as above. The constant term in \eqref{eqn:detfinal} is due to the identity element contribution and  given in appendix~\ref{app:determinant}.

\subsection{JT examples} \label{sec:2.3examples}
Let us briefly explain how these determinants and resolvents arise for JT gravity coupled to matter. Consider a massive scalar field minimally coupled to JT gravity. The path integral is given, up to boundary terms, by
\be
Z_{\t{JT + matter}} = \int \mathcal{D} g \mathcal{D} \Phi \exp\lr{\frac{1}{2} \int \sqrt{g}\phi \lr{R+2}} \hspace{.05cm}Z_{\t{matter}}[g],
\ee
where the matter partition function is
\be \label{eqn:matterdet}
Z_{\t{matter}}[g] = \int \mathcal{D} \phi e^{-S[\phi,g]} = \frac{1}{\sqrt{\det \lr{\Delta+m^2}}}, \qquad S[\phi,g] = \frac{1}{2}\int_\Sigma \sqrt{g} \lr{\phi \Delta \phi + m^2 \phi^2}.
\ee
where we define the mass through $m^2=s(s-1)$, with $s\geq 1$  the scaling dimension of the boundary operator dual to $\phi$. We see the matter partition function reduces down to a determinant. 
After integrating out the dilaton we localize onto the moduli space of constant negative curvature geometries,\footnote{By appropriately deforming the JT gravity action we can also localize onto the moduli space of constant negative curvature surfaces with conical points which we will describe in detail in section \ref{sec:3}.} and so is given by the determinant calculated in section \ref{sec:2.2dets}. Another interesting observable when matter is present is the two-point function/resolvent $\langle \phi(z) \phi(w) \rangle$ given by $R_{\Sigma}(s;z,w)$. On the upper half plane the scalar Laplacian is given by
\be
ds^2 = \frac{dx^2 + dy^2}{y^2}, \qquad \Delta = -y^2 \lr{\partial_x^2 + \partial_y^2}.
\ee
Defining the complex coordinates $z=x+iy$, the resolvent is given by
\be \label{eqn:2ptfn_UHP}
    R_{\mathbb{H}}(s;z,w)=\frac{\Gamma(s)^2}{4\pi \Gamma(2s) } \sech^{2s} \lr{\frac{\ell(z,w)}{2}}\hspace{.05cm} {}_2F_1 \lr{s,s,2s;\sech^2 \lr{\frac{\ell(z,w)}{2}}}.
\ee
where $\ell(z,w)$ is the geodesic distance between $z$ and $w$. 
In JT gravity we want to dress the two-point function to the boundary schwarzian. Since the geodesic distance goes to infinity near the boundary the two-point function simplifies to
\be \label{eqn:2ptfn_bdy_qft}
R_{\mathbb{H}}(z,w) \underrel{\ell \to \infty}{=} \frac{4^s \Gamma(s)^2}{4\pi \Gamma(2s)} e^{-s \ell(z,w)} \lr{1 + \mathcal{O}\lr{e^{-\ell}}}\,,
\ee
and so the geodesic approximation becomes exact. On an arbitrary surface $\Sigma$ the two-point function is obtained by summing over the group in equation \eqref{eqn:imageresolvent} so that we obtain
\be \label{eqn:resolvent_sum_images}
\begin{aligned}
R_\Sigma(s;z,w) &= \frac{\Gamma(s)^2}{4\pi \Gamma(2s) } \sum_{T \in \Gamma} \sech^{2s} \lr{\frac{\ell(z,T\cdot w)}{2}}\hspace{.05cm} {}_2F_1 \lr{s,s,2s;\sech^2 \lr{\frac{\ell(z,T\cdot w)}{2}}}\\
  & \hskip -6pt \underrel{\ell \to \infty}{=} \frac{4^s \Gamma(s)^2}{4\pi \Gamma(2s)} \sum_{T\in\Gamma} e^{-s \ell(z,T\cdot w)} \lr{1 + \mathcal{O}\lr{e^{-\ell}}}.
\end{aligned}
\ee
This implements a summation over all possible geodesics connecting the points $z,w$. To dress the operator insertions to the boundary fluctuations we parameterize the boundary by the curve $(x,y)=(F(\tau),\epsilon F'(\tau))$ where $\tau$ is an affine time, $\epsilon$ is the boundary cutoff, and we have enforced the condition that the induced metric on the boundary is $\sqrt{h}=1/\epsilon$. We consider inserting operators at two boundary points labelled by times $\tau_1, \tau_2$. The complex coordinates at these times are $z=F(\tau_1)+i \epsilon F'(\tau_1)$ and $w=F(\tau_2)+i \epsilon F'(\tau_2)$. The geodesic distance between these points is given by 
\be
\ell(z,w) =\log \lr{ \frac{1}{\epsilon^2}\frac{\lr{F(\tau_1)-F(\tau_2)}^2}{F'(\tau_1) F'(\tau_2)} + \mathcal{O}\lr{\frac{1}{\epsilon}}}\,.
\ee
We will define the two-point function dressed to the boundary fluctuations by $G(\tau_1,\tau_2)$, and subtract off the usual divergences to obtain
\begin{equation}
    G(\tau_1,\tau_2)=\lim_{\epsilon\rightarrow 0} \frac{4\pi \Gamma(2s)}{4^s \Gamma(s)^2 \epsilon^{2s}}  R_{\mathbb{H}}(s;z,w) = \left(\frac{F'(\tau_1)F'(\tau_2)}{\lr{F(\tau_1)-F(\tau_2)}^2}\right)^{s}\,.
\end{equation}
This is the standard result for the matter two-point function dressed to the Schwarzian boundary at the level of the disk. In terms of the thermal circle reparameterization $f(\tau)$ defined by $F(\tau) \equiv \tan \frac{\pi}{\beta} f(\tau)$ the regularized resolvent is
\begin{equation} \label{eqn:Gdisk}
G(\tau_1,\tau_2)=\left(\frac{f'(\tau_1)f'(\tau_2)}{\frac{\beta^2}{\pi^2}\sin^2\frac{\pi}{\beta}\lr{f(\tau_1)-f(\tau_2)}}\right)^{s}\,.
\end{equation}
On a more complicated surface obtained from a quotient by $\Gamma$ there are additional geodesics connecting the two boundary points. Since $w\cong T\cdot w$ for $T \in \Gamma$ the geodesic connecting $z$ to $T \cdot w$ on the disk becomes a distinct geodesic connecting $z$ to $w$ on the quotient geometry. We can find the geodesic distance between the point $z$ and $T \cdot w$ to be
\be
\ell\lr{z,T \cdot w} = \log \lr{\frac{1}{\epsilon^2} \frac{ \lr{F(\tau_1)-T \cdot F(\tau_2)}^2 }{F'(\tau_1)\lr{T \cdot F(\tau_2)}'} + \mathcal{O}\lr{\frac{1}{\epsilon}}}, \qquad T \cdot F(\tau) = \frac{a F(\tau) + b}{c F(\tau) + d},
\ee
The boundary dressed two-point function on a quotient surface can then be immediately obtained from equation \eqref{eqn:resolvent_sum_images}
\be \label{2pt_final}
G_{\Sigma}(\tau_1, \tau_2) = \lim_{\epsilon\rightarrow 0} \frac{4\pi \Gamma(2s)}{4^s \Gamma(s)^2 \epsilon^{2s}}  \sum_{T \in \Gamma} e^{-s \ell(z,T\cdot w)} =\sum_{T \in \Gamma} \left(\frac{F'(\tau_1)(T\cdot F(\tau_2))'}{\lr{F(\tau_1)-T \cdot F(\tau_2)}^2}\right)^{s}.
\ee
This formula includes all possible geodesics on the surface $\Sigma$ connecting the specified boundary points, and different group elements $T$ lead to  geodesics with different windings/self-intersections on $\Sigma$.
This includes geodesics that self-intersect any number of times. However, we are still left with the challenge of integrating over the Schwarzian and performing the sum over the group elements. We now explain some simple examples. 

\subsubsection{Double trumpet}
The metric for the double trumpet can be written as
\be
ds^2 = dr^2 + \cosh^2\lr{r} d \theta^2, \qquad \theta \cong \theta + b.
\ee
The two asymptotic boundaries are at $r \to \pm \infty$, and the only primitive geodesic of length $b$ is at $r=0$. 
The double trumpet is obtained by taking the quotient of the upper half-plane $\mathbb{H}/\Gamma$ by the group $\Gamma = \lrb{T_b}$ generated by the element
\be
T_b = \begin{pmatrix} \exp\lr{\frac{b}{2}} & 0\\0 &  \exp\lr{-\frac{b}{2}}\end{pmatrix},
\ee
which identify points in the UHP plane by $z \cong e^{b} z$. In the fundamental domain the geodesic throat of length $b$ goes along the imaginary axis from $z=i$ to $z=i e^b$. The group elements are hyperbolic and take the form $T_b^m$ for $m \in \mathbb{Z}$, and correspond to the geodesics that winds $m$ times around the throat. The resolvent is given by
\be
R_{\mathbb{H}/\langle T_{b} \rangle }(s;z,w)= \sum_{m} R_{\mathbb{H}}(s;T_{b}\cdot z,w)\,.
\ee
The resolvent trace and determinant have been discussed in section~\ref{sec:2.2dets} and we simply repeat the result
\begin{equation}
    {\rm tr}R_{\mathbb{H}/\langle T_{b} \rangle }= \frac{2b}{2s-1}\sum_{m=1}^{\infty} \frac{e^{-smb}}{1-e^{-mb}}, \qquad \det\lr{{\Delta}+s(s-1)}_{\t{DT}} = \prod_{m=0}^\infty \lr{1-e^{-\lr{s+m}b}}^2\,.
\end{equation}
where the quantity in parenthesis is squared since there are two orientations for the primitive geodesic, and the determinant should be understood to be up to multiplicative constants.

\paragraph*{One sided two-point function.}
We first consider the correlator where we insert two matter operators $\lrb{\phi_L(\tau_1)\phi_R(\tau_2)}$ onto the same asymptotic boundary and dress them to the boundary fluctuations. We take the operator to be located at $z=F_R(\tau_1) + i \epsilon F_R'(\tau_1)$ and $w=F_R(\tau_2) + i \epsilon F_R'(\tau_2)$. The quotient $z \cong T \cdot z$ enforces a periodicity constraint on the boundary\footnote{This is different from the $\tanh$ one in the JT review, but equivalent The Schwarzian derivative itself is $\operatorname{PSL}(2, \mathbb{R})$ invariant: if $F=\frac{a G+b}{c G+d}$, then $\{F, t\}=\{G, t\}$.} 
\be
F_R(\tau+\beta) = T_b \cdot F_R(\tau), \qquad T_b^m \cdot F(\tau) = e^{m b} F(\tau).
\ee
We can introduce the parameterization $F_R(\tau) = \exp\lr{\frac{b}{\beta} f_R(\tau)}$ with $f_R(\tau+\beta) = f_R(\tau)+\beta_R$ which respects this identification. We can immediately apply \eqref{2pt_final} to find the two-point function
\begin{equation} \label{eqn:onesided_DT}
    G^{\t{RR}}_{\t{DT}}(\tau_1,\tau_2)
    =\sum_{m=-\infty}^{\infty} \left(\frac{F_R'(\tau_1)(T_b^{m}\cdot F_R(\tau_2))'}{(F_R(\tau_1)-T_b^{m}\cdot F_R(\tau_2))^2}\right)^{s}=
   \sum_{m=-\infty}^{\infty}
    \left(\frac{f_{R}'(\tau_1)f_{R}'(\tau_2)}{\frac{4\beta_R^2}{b^2}\sinh^2\left[\frac{b}{2 \beta_R}
    \left(f_{R}(\tau_1)- f_{R}(\tau_2+m\beta_R)\right)\right]}\right)^{s}  \,.
\end{equation}
We have used that the summation over the group $\Gamma$ is the same as a summation over powers of $T_b^m$. The sum over $m$ is over windings of the geodesic around the double trumpet as it connects the two points. Positive and negative integers indicate which direction the geodesic winds around the trumpet. The largest contribution is given by the shortest geodesic which doesn't wind and is given by $m=0$. This reproduces the one sided two-point function on the double trumpet computed in \cite{Mertens:2019defects}. 

\begin{figure}
    \centering
    \includegraphics[width=11cm]{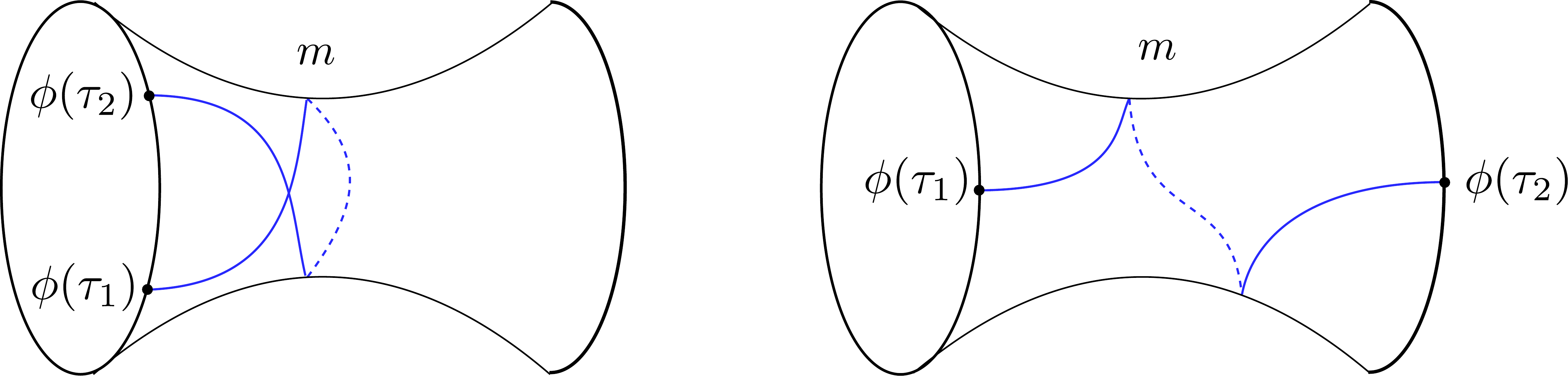}
    \caption{Two-point function on the double trumpet. We have geodesics connecting the two operator insertions that wind an integer $m$ times around the trumpet.}
\end{figure}

\paragraph*{Two sided two-point function.} 
The correlator with matter insertions on opposite boundaries $\lrb{\phi_L(\tau_1)\phi_R(\tau_2)}$ also immediately follows from the previous result. The operators are now dressed to independent boundary wiggles $z=F_L(\tau_1) + i \epsilon F_L'(\tau_1)$ and $w=F_R(\tau_2) + i \epsilon F_R'(\tau_2)$. The quotient enforces the periodicity constraint on both boundaries $F_{L,R}(\tau+\beta_{L,R})=T_b \cdot F_{L,R}(\tau)$, and we can define $F_{L,R}(\tau) = \mp \exp\lr{\frac{b}{\beta_{L,R}} f_{L,R}(\tau)}$ which satisfies the constraint as long as $f_{L,R}(\tau+\beta_{L,R}) = f_{L,R}(\tau)+\beta_{L,R}$ where the two boundaries can have independent temperatures $\beta_{L,R}$. The relative minus sign is because the left boundary is at $\Re(z)<0$, so we must have that $F_L < 0$ whereas $F_R > 0$. Performing the sum over the group \eqref{2pt_final} as in the one sided case we obtain
\begin{equation}
    G^{\t{LR}}_{\t{DT}}(\tau_1,\tau_2)
    =\sum_{m=-\infty}^{\infty} \left(\frac{F_L'(\tau_1)(T_b^{m}\cdot F_R(\tau_2))'}{(F_L(\tau_1)-T_b^{m}\cdot F_R(\tau_2))^2}\right)^{s}=
   \sum_{m=-\infty}^{\infty}
    \left(\frac{f_{L}'(\tau_1)f_{R}'(\tau_2)}{\frac{4 \beta_L \beta_R }{b^2}\cosh^2\left[\frac{b}{2}
    \left( \frac{f_{L}(\tau_1)}{\beta_L}- \frac{f_{R}(\tau_2+m\beta_R)}{\beta_R}\right)\right]}\right)^{s}  \,.
\end{equation}
For the two-sided correlator the geodesic can also wind $m$ times around the trumpet. The dominant contribution is given by the geodesic with smallest length, which is given by $m=0$ winding. 

\subsubsection{Conical defect}
\begin{figure}
    \centering
    \includegraphics[width=3.6cm]{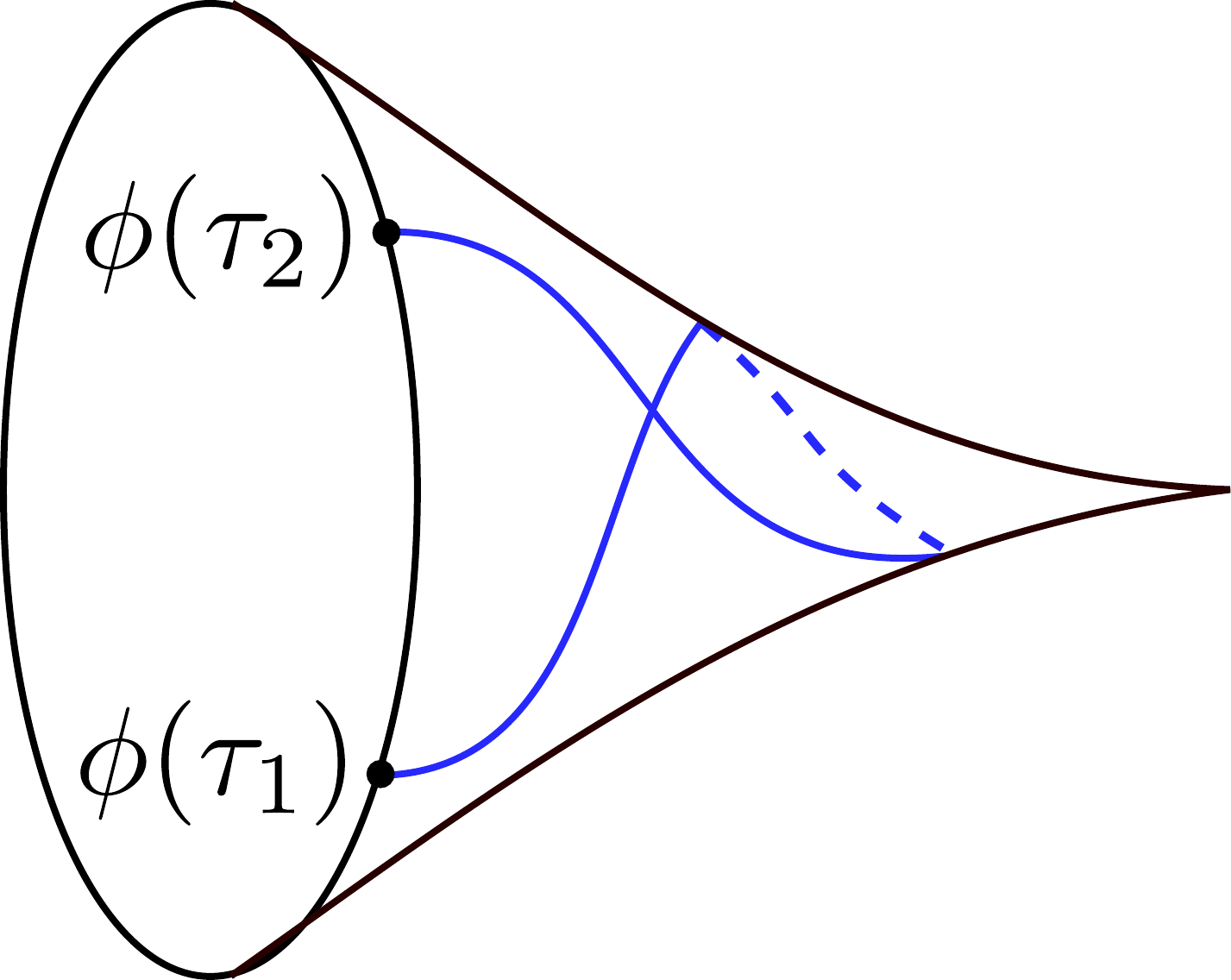}
    \caption{Two-point function on the conical defect geometry. The geodesics connecting the two operators can wind up to $n-1$ times around the defect.}
\end{figure}
We now consider the geometry with a single conical defect of opening angle $\theta = \frac{2\pi}{n}$ by taking the quotient by the group $\Gamma=\langle T_{\theta}\rangle$. The generator is given by
\be
T_\theta=\left(\begin{array}{ll}
\cos \frac{\pi}{n} & -\sin \frac{\pi}{n} \\
\sin \frac{\pi}{n} & ~~\cos \frac{\pi}{n}
\end{array}\right)\,,
\ee
and the conical defect is located at the fixed point $z=i$. This group has $n$ elements corresponding to powers of $T_\theta^m$ for $m=0,\ldots,n-1$. The determinant on this geometry is not very interesting since the cone has no closed geodesics. Instead we will consider the two-point function dressed to the boundary schwarzian. As in the case of the disk, we insert the two operators at $z=F(\tau_1) + i \epsilon F'(\tau_1)$ and $w=F(\tau_2) + i \epsilon F'(\tau_2)$. The identification $z \cong T_\theta \cdot z$ enforces the condition
\be
F(\tau+\beta) = T_\theta \cdot F(\tau). 
\ee
We can also use the parameterization $F(\tau) = \tan \frac{\theta}{2 \beta} f(\tau)$, with $f(\tau + \beta) = f(\tau) + \beta$. Applying the two-point function formula \eqref{2pt_final} we have 
\be
G_{\text{Defect}}(\tau_1,\tau_2) =\sum_{m=0}^{n-1} \left(\frac{F'(\tau_1)(T_{\theta}^{m}\cdot F(\tau_2))'}{\lr{F(\tau_1)-T_{\theta}^{m}\cdot F(\tau_2)}^2}\right)^{s}
=\sum_{m=0}^{n-1}\left(\frac{f'(\tau_1)f'(\tau_2)}{\frac{4 \beta^2}{ \theta^2}\sin^2\frac{\theta}{2\beta}\left[f(\tau_1+m \beta)-f(\tau_2)\right]}\right)^{s}\, ,
\ee
There are $n$ geodesics connecting the two operator insertions on the boundary. These geodesics wind around the defect up to $n-1$ times which is the summation over $m$. Note that for $n=1$ the opening angle becomes $\theta=2\pi$ and we get the answer for the disk. The above is not the complete answer, since we must still integrate over the boundary fluctuations.

If we consider the case of a massless field $s=1$ we can perform the summation over winding geodesics to find 
\be
G^{s=1}_{\text{Defect}}(\tau_1,\tau_2) = \frac{f'(\tau_1)f'(\tau_2)}{\frac{\beta^2}{\pi^2} \sin^2 \frac{\pi}{\beta}\left[f(\tau_1)-f(\tau_2)\right]},
\ee
which is precisely the two-point function for a massless field on the disk \eqref{eqn:Gdisk}. This is the expected answer for a conformal scalar since the correlator on the defect geometry is related to the correlator on the disk through a conformal rescaling. In this case the rescaling is trivial at the boundary so the answers agree. We see that including all of the winding geodesics is crucial to recover the correct properties of the matter fields.\footnote{We still have to integrate over the schwarzian mode. On the defect geometry the $n=2$ boundary modes are no longer zero modes and must be included in the integral. Even though we have reduced the two point function to the standard disk answer, we must still include these boundary modes when computing the two-point function.}

\subsubsection{Two conical defects}
To get a disk with two conical defects we need to generate a Fuchsian group with two elliptic generators $\Gamma = \lb T_{\theta_1}, T_{\theta_1} \rb$. We choose the generators to be
\begin{figure}
    \centering
    \includegraphics[width=9cm]{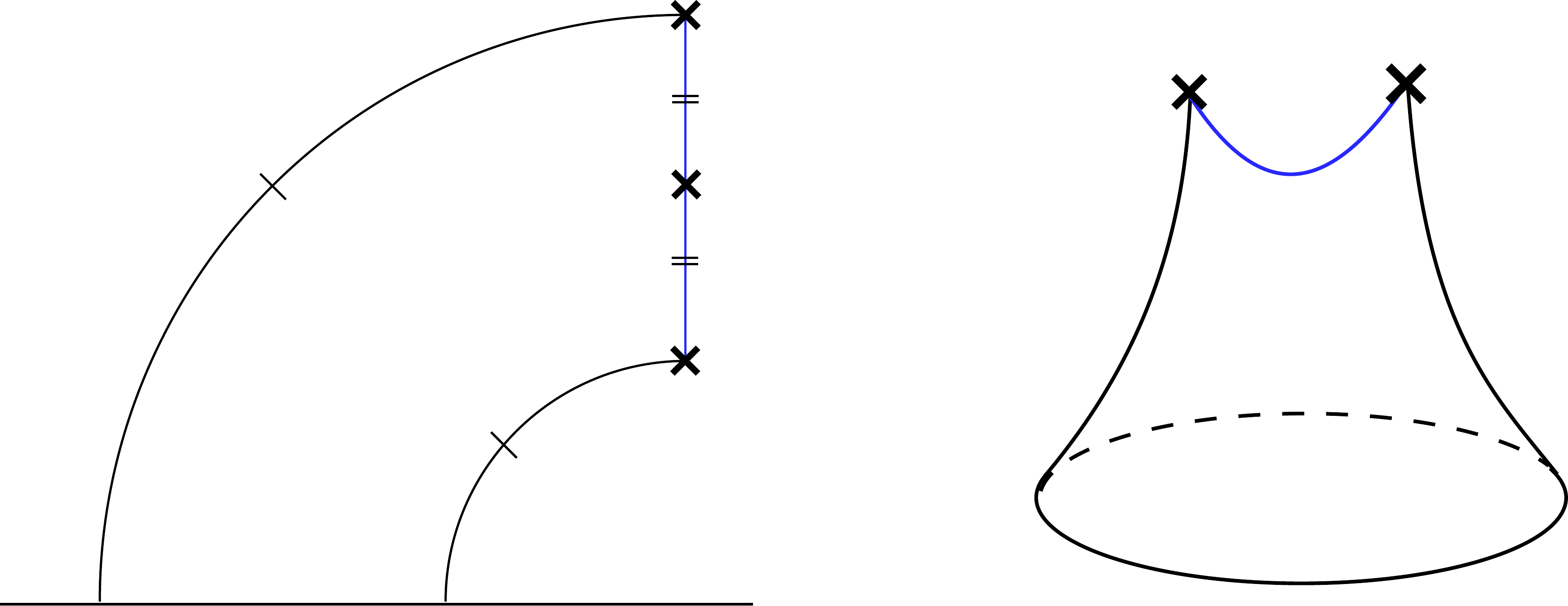}
    \caption{The disk with two defects can be represented in the UHP by the pictured fundamental domain, with the two defects represented by ``\textbf{x}''. The single primitive closed geodesic is in blue, and travels between the defects. We also have any number of windings of this geodesic.}
\end{figure}
\begin{equation}
    T_{\theta_1}=\begin{pmatrix}
        \cos\frac{\theta_1}{2} & e^{-\ell/2}\sin\frac{\theta_1}{2}  \\
        e^{\ell/2} \sin\frac{\theta_1}{2} & \cos\frac{\theta_1}{2} 
    \end{pmatrix}\,, \quad T_{\theta_2}=\begin{pmatrix}
        \cos\frac{\theta_2}{2} & e^{\ell/2} \sin\frac{\theta_2}{2} \\
       e^{-\ell/2}\sin\frac{\theta_2}{2} & \cos\frac{\theta_2}{2} 
    \end{pmatrix}\,.
\end{equation}
In the UHP the fixed point of $T_{\theta_1}$ is at $z=ie^{-\ell/2}$ while the fixed point of $T_{\theta_2}$ is at $z=ie^{\ell/2}$, which are the locations of the conical defects with opening angles $\theta_1, \theta_2$ respectively. The simplest example, which captures many features of multi defect surfaces, is to consider the case where both of the angles take the value $\theta_{1,2}=\pi$. In this case the simplest group products take the form 
\be
T_{\theta_1}^2=T_{\theta_2}^2=\mathbb{I},\qquad  T_{2\ell}\equiv T_{\theta_2 }T_{\theta_1}=\begin{pmatrix}
        e^{\ell} & 0 \\
       0 & e^{-\ell}
    \end{pmatrix},
\ee
where we see that a product of two elliptic elements becomes a hyperbolic element $T_{2\ell}$. There are thus three primitive elements in the group: two elliptic $T_{\theta_1}, T_{\theta_2}$ and one hyperbolic $T_{2\ell}$. From the general formula for the determinant \eqref{eqn:detfinal} we immediately get\footnote{The careful  reader may wonder why the Selberg zeta function does not get a square like the double trumpet case. The reason is that the closed geodesic in this case is not orientable. The closed geodesic with total length $2\ell$ is actually obtained by gluing a copy of a geodesic segement with length $\ell$ back to itself in the opposite direction.}
\begin{equation}
    \det (\Delta+s(s-1)) = \underbrace{\prod_{m=0}^{\infty} \lr{1-e^{-2\ell (s+m)}}}_{\t{closed geodesic}} \underbrace{\frac{\Gamma\left(s+\frac{1}{2}\right)}{\Gamma\left(s\right)}.}_{\t{defect contribution}}
\end{equation}
The Selberg zeta function contribution is given by the single geodesic traversing between the two defects, which has length $2\ell$. This originates from the single primitive hyperbolic element $T_{2\ell}$. We have also included the contribution coming from the elliptic elements in the resolvent, which correspond to the two defects. One interesting feature of this surface is that the hyperbolic element is generated by products of elliptic elements, and that the resulting closed geodesic touches the conical defects. 

To calculate the two-point function \eqref{2pt_final} we must sum over the group. By inspection we can find that all elements of the group take the form
\begin{equation}
T_{2\ell}^m =  
    \begin{pmatrix}
        e^{m \ell} & 0  \\
       0 & e^{-m \ell} 
    \end{pmatrix}\,, \quad     
        T_{\theta_1} T_{2\ell}^m = 
        \begin{pmatrix}
        0 & e^{\lr{m+\frac{1}{2}} \ell}  \\
       e^{-\lr{m+\frac{1}{2}} \ell} & 0 
    \end{pmatrix}\,,
\end{equation}
with $m$ an integer. From the fundamental domain of the schwarzian we see we must identify $F(\tau+\beta)= e^{2 \ell} F(\tau)$, which can be expanded in terms of the identification $F(\tau) = \exp \lr{\frac{2 \ell}{\beta} f(\tau)}$. We can immediately write the answer for the resolvent to be
\begin{align}
&G_{\text{Defects}}(\tau_1,\tau_2) =\sum_{m=-\infty}^{\infty} \left(\frac{F'(\tau_1)(T_{2\ell}^{m}\cdot F(\tau_2))'}{\lr{F(\tau_1)-T_{\theta}^{m}\cdot F(\tau_2)}^2}\right)^{s} + \left(\frac{F'(\tau_1)(T_{\theta_1}T_{2\ell}^{m}\cdot F(\tau_2))'}{\lr{F(\tau_1)-T_{\theta}^{m}\cdot F(\tau_2)}^2}\right)^{s} \nn
\\
&=\sum_{m=-\infty}^{\infty} \left(\frac{f'(\tau_1)f'(\tau_2)}{\frac{\beta^2}{ b^2}\sinh^2\left[\frac{b}{\beta}
    \left(f(\tau_1)- f(\tau_2+m\beta)\right)\right]}\right)^{s} +\left(\frac{f'(\tau_1)f'(\tau_2)}{\frac{\beta^2}{ b^2}\sinh^2\left[\frac{b}{\beta}
    \left(f(\tau_1)+ f(\tau_2+m\beta)-\frac{\beta}{2}\right)\right]}\right)^{s}  \, .
\end{align}
The first term corresponds to geodesics that wind around the throat, whereas the second term are geodesics that wind around the defects. 

The above example was for the special case where both defects have deficit angle $\pi$, but the construction immediately extends to more general opening angles smaller than $\pi$. We briefly explain the main differences.
\begin{itemize}[topsep=3pt,itemsep=-1ex,partopsep=1ex,parsep=1ex]
    \item To ensure the quotient surface  has one asymptotic boundary and two defects, the parameter $\ell$ must satisfy
\begin{equation}
    e^{\ell}>\frac{\sin \theta_1\left(1+\cos \theta_2\right)}{\sin \theta_2\left(1-\cos \theta_1\right)}\,.
\end{equation}
Geometrically, it means that two sharp defects can not be arbitrarily close to each other.

    \item There are infinitely many closed geodesics on the surface, generated by the primitive ``words" of $T_{\theta_{1,2}}$ such as $T_{\theta_1}^2, T_{\theta_1} T_{\theta_2}, T_{\theta_2}^2 T_{\theta_1}^7, \ldots$ as long as the ``word'' is hyperbolic.

    \item Such geodesics generically self-intersect. For example, $T_{\theta_1}T_{\theta_2}^{-1}$ is a geodesic winding around both defects once with a self-intersection. 
\end{itemize}

\subsubsection{The handle disk}

\begin{figure}
\centering
\begin{subfigure}
         \centering
         \includegraphics[width=15cm]{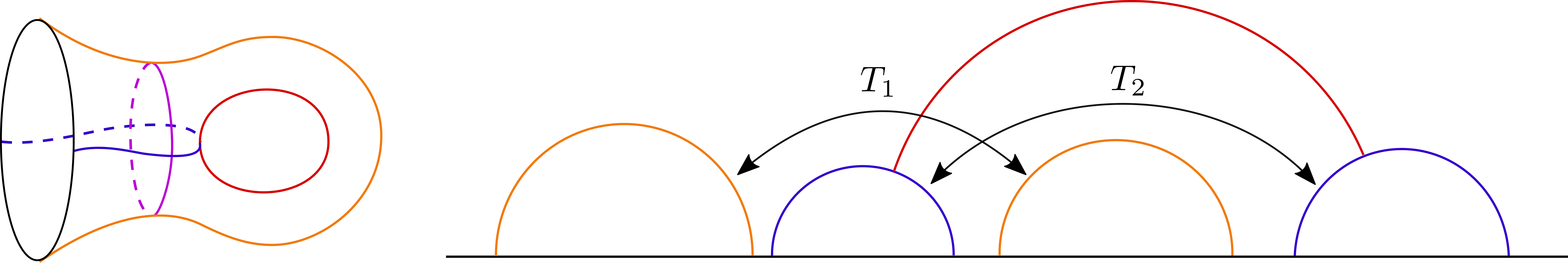}
    \end{subfigure}
     \caption{The handle disk can be constructed by identifying the orange and blue geodesic semi-circles in the UHP. The fundamental domain is the exterior of all the semi-circles. The two identifications are the generators of the Fuchsian group that generate the surface. The red and purple curves are closed geodesics on the surface. In the right figure we omit the purple closed geodesic because it is broken into four segments with the pictured fundamental domain.}
     \label{fig:handledisk}
\end{figure}

We now consider the handle disk which has a non-trivial topology. To build the surface we can cut the handle disk along two geodesics, and flatten it onto the UHP with the identification in figure~\ref{fig:handledisk}.
Therefore, the Fuchsian group is generated by two hyperbolic elements that identify the associated geodesics: the first semicircle with the third and the second semicircle with the fourth.\footnote{Fuchsian groups generated by identifying pairs of semicircles are called the Schottky groups.} 
The generators will be denoted by $T_1$ and $T_2$,\footnote{The form of the hyperbolic generators is given as follows. A hyperbolic element that identifies semi-circles centered along the real axis at points $c_{1,2}$ with radii $r_{1,2}$ is
\begin{equation*}
   T(c_1,c_2;r_1,r_2)= \left(r_1 r_2 (c_{1 2}-r_{1 2})(c_{1 2}+r_{1 2})\right)^{-\frac{1}{2}}  \begin{pmatrix}
        -c_2 c_{1 2} + r_2 r_{1 2} & (c_1 c_2 + r_1 r_2)(c_{1 2} - r_{1 2}) \\
       - c_{1 2} & c_1 c_{1 2} - r_1 r_{1 2}
    \end{pmatrix}
\end{equation*}
where $c_{1 2} =|c_1-c_2|$ and $r_{1 2} =|r_1-r_2|$.}
and the fundamental domain is to remove the four half disks from $\mathbb{H}$. 

As in the case of the two-defect surface any hyperbolic word $T_1^{k_1}T_2^{k_2}T_1^{k_3}\ldots\,$, on the condition that it is primitive, corresponds to a primitive closed geodesic on the handle disk. A generic hyperbolic word corresponds to an integer winding of a primitive geodesic. As an example, the dark red geodesic in figure~\ref{fig:handledisk} is associated to $T_2$ while the purple one is the closed geodesic separating the torus with the geodesic boundary, generated by $T_1 T_2 T_1^{-1} T_2^{-1}$. More complicated closed geodesics typically have self-intersections. Each closed geodesics (with an associated hyperbolic word) contributes to the determinant, and so it is difficult to write down a simple expression for the full determinant. 

\begin{figure}
\centering
\includegraphics[width=14cm]{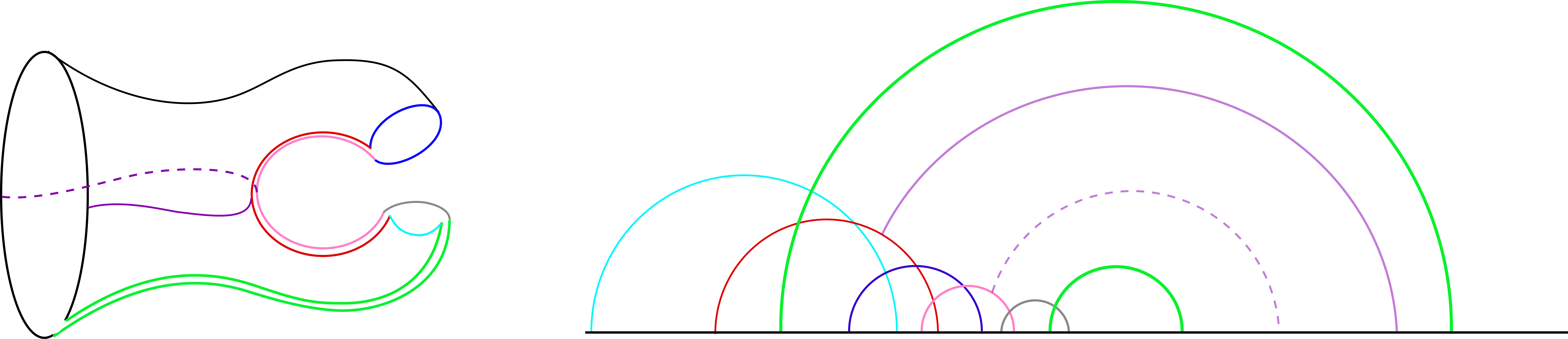}
    \caption{An alternative representation of the handle disk fundamental domain. All the semi-circles are geodesics, and fundamental domain is the bounded region inside all the semi-circles. The identifications are red/pink, blue/light blue/gray, green/green. The purple geodesic winds around the handle. In this representation the asymptotic boundary is in one connected segment. The two green semi-circles are mapped to each other by the group element $\tilde T_1 \tilde T_2 \tilde T_1^{-1} \tilde T_2^{-1}.$}
    \label{fig:handledisk2}
\end{figure}

For the two-point function we must consider boundary anchored geodesics such as the orange and the blue geodesics in figure~\ref{fig:handledisk}. It is convenient to use a different fundamental domain for the handle disk where the asymptotic boundary is in one connected segment of the UHP. 
To achieve this, one can first diagonalize $T_1 T_2 T_1^{-1} T_{2}^{-1}$, which is associated to the geodesic separating the asymptotic boundary from the handle, to be $-\t{diag}(e^{\ell/2},e^{-\ell/2})$. 
This modifies the generators of the group to be $\tilde{T}_1, \tilde{T}_2$ with generated group $\tilde{\Gamma}$.\footnote{To get $\tilde{T}_{1,2}$ and  $\tilde{\Gamma}$, one should conjugate every element in $\Gamma$ with a proper SL(2,$\mathbb{R}$) matrix $V$ so that $V \cdot  T_1 T_2 T_1^{-1} T_{2}^{-1}\cdot V^{-1}=-\t{diag}(e^{\ell/2},e^{-\ell/2})$. Then $\tilde{\Gamma}$ is $V\cdot \Gamma \cdot V^{-1}$, and the generators are $\Tilde{T}_i = V \cdot T_i V^{-1}$ for which an explicit although complicated form can be obtained.} The resulting fundamental domain is pictured in figure~\ref{fig:handledisk2}. The fundamental domain looks very similar to the double trumpet, and we can parameterize the Schwarzian boundary by $z=F(\tau)+i\epsilon F'(\tau)$ with $F(\tau)=\exp\left(\frac{\ell}{\beta}f(\tau)\right)$. The two point function is given by
\begin{equation}\label{eqn:handledisk2pt}
    G(\tau_1,\tau_2) = \sum_{\gamma \in \tilde{\Gamma}}\left(\frac{F'(\tau_1) \lr{\gamma \cdot F(\tau_2)}' }{\left( F(\tau_1)-\gamma\cdot F(\tau_2)\right)^2}\right)^s\,.
\end{equation}
Note that the parameterization is the same as the one-sided two-point function on the double trumpet \eqref{eqn:onesided_DT}, but the handle disk case is much more complicated due to its group $\tilde{\Gamma}$. There are distinct classes of geodesics on the handle disk, and we now explain how the sum over geodesics can be categorized into equivalence classes using large diffeomorphisms.

The moduli space of the handle disk is composed of schwarzian fluctuation and the moduli of the torus with geodesic boundary denoted by $\mathcal{M}_{1,1}$. The moduli space is obtained from Teichmuller space $\mathcal{T}_{1,1}$ after modding out by the action of large diffs, also known as the mapping class group (MCG). The MCG is generated by three elements $\{\sigma, P, U\}$ and has a simple action on the generators of the Fuchsian group \cite{wolpert1983kahler}
\begin{equation}
    \begin{array}{ll}
\sigma(\tilde T_1)=\tilde T_1^{-1}, & P(\tilde T_1)= \tilde T_2, \\
\sigma(\tilde T_2)=\tilde T_2, & P(\tilde T_2)=\tilde T_1,
\end{array} \  \begin{aligned}
& \tilde U(T_1)=\tilde T_1 \tilde T_2, \\
& U(\tilde T_2)=\tilde T_2\,.
\end{aligned}
\end{equation}
The MCG acts on a group element built from products of the generators by acting on each generator independently. This action can be understood as a map between cycles of the surface. Since each cycle has an associated closed geodesic, we can think of this action as generating a map between closed geodesics on the surface. When considering boundary anchored geodesics, this becomes a map between boundary anchored geodesics. As an example, it can be checked that the MCG preserves the group element $\tilde T_1 \tilde T_2 \tilde T_1^{-1} \tilde T_2^{-1}$ up to conjugation,\footnote{For example, $\sigma\left(\tilde T_1 \tilde T_2 \tilde T_1^{-1} \tilde T_2^{-1}\right)$ is $\tilde T_1^{-1} \tilde T_2 \tilde T_1 \tilde T_2^{-1}$, which is conjugate to $\left(\tilde T_1 \tilde T_2 \tilde T_1^{-1} \tilde T_2^{-1}\right)^{-1}$ by $\tilde T_1$.} so the geodesic splitting the handle from the asymptotic boundary is fixed. We can consider a group element $\gamma\in \tilde \Gamma$ corresponding to a particular boundary anchored geodesic and act on it with the MCG to generate the equivalence class $\{\gamma\}_{\rm MCG}$.\footnote{The geodesic corresponding to $\gamma$ is the geodesic connecting the boundary points $F(\tau_1)$ and $\gamma \cdot F(\tau_2)$ if we unwrap the handle disk on the UHP.} As an example the element $\tilde T_1$ corresponds to the purple boundary anchored geodesics that winds the handle in figure~\ref{fig:handledisk2}. We can generate all other geodesics that wind the handle and do not self-intersect by acting with the MCG
\begin{equation}
    \t{non self-intersecting geodesics: }\{\tilde T_1\}_{\rm MCG}=\{\tilde T_1,\tilde T_2, \tilde T_1 \tilde T_2,\tilde T_1^2 \tilde T_2,\tilde T_1^3 \tilde T_2, \tilde T_1^2 \tilde T_2^3, \ldots\}\,.
\end{equation}
The summation over the elements $\{\tilde T_1\}_{\rm MCG}$ in the two point function \eqref{eqn:onesided_DT} implements a sum over all non-self intersecting geodesics that wind the handle. These contributions were resumed in \cite{Saad:2019pqd} and proved to give the ramp behavior for the late time two-point function.

This method can also be used to partially classify self-intersecting geodesics that wind around the handle. A complete closed form classification appears to be highly non-trivial, but for one self-intersection we can find explicit examples of simple equivalence classes. For boundary anchored once self-intersecting geodesics that wind the handle, two equivalence classes are given by starting with base elements $\tilde T_1^2 \tilde T_2^2$, $\tilde T_1 \tilde T_2\tilde T_1\tilde T_2\tilde T_1^{-1}\tilde T_2^{-1}$ and acting with the MCG
\be \label{eqn:equivclass2}
\begin{aligned}
   \t{one self-intersection: } &\{\tilde T_1^2\tilde T_2^2\}_{\rm MCG}=\{\tilde T_1^2\tilde T_2^2,\tilde T_1\tilde T_2\tilde T_1 \tilde T_2^{-1},\tilde T_1^3 \tilde T_2 \tilde T_1 \tilde T_2,  \ldots\}, \\
   & \{\tilde T_1 \tilde T_2 \tilde T_1 \tilde T_2 \tilde T_1^{-1} \tilde T_2^{-1}\}_{\rm MCG}=\{\tilde T_1 \tilde T_2 \tilde T_1 \tilde T_2 \tilde T_1^{-1} \tilde T_2^{-1}, \tilde T_1 \tilde T_2 \tilde T_1^3 \tilde T_2 \tilde T_1^{-1} \tilde T_2^{-1}, \ldots\}\,. 
\end{aligned}
\ee
Note that the above is not a complete classification of all once self-intersecting geodesics, but we believe the first equivalence class will always contain the shortest self-intersecting geodesics on the surface.\footnote{The argument is that the length of self-intersecting geodesics increases with additional windings, and the number of windings is controlled by the word-length of the generator.} We can write down other equivalence classes systematically using the group theory description.\footnote{Note that geodesics with more self-intersections are generated by words with at least six generators. The greater length of the words implies a greater length of the corresponding geodesic.} It would be interesting to perform the integral over the boundary fluctuations in \eqref{eqn:handledisk2pt} for the self-intersecting geodesics falling into the equivalence classes \eqref{eqn:equivclass2} to confirm that they do not contribute to the late time two-point function \cite{Saad:2019pqd}.

\section{JT gravity path integral} \label{sec:3}

\subsection{Gauge fixing the path integral}
We now explain how to gauge fix the gravity path integral for JT gravity. For the case of compact surfaces see \cite{Saad:2019lba}, while for non-compact surfaces also see \cite{Moitra:2021uiv,Charles:2019tiu,Choi:2021nnq,Choi:2023syx}. The integral is defined by
\be \label{Z_JT}
Z = \int \frac{\mathcal{D} g \mathcal{D} \Phi}{\t{V}(\t{Diff})} e^{-I_{\t JT}[g,\Phi]},
\ee
where we divide out by the volume of diffeomorphisms. The JT gravity action is given by
\be \label{I_JT}
I_{\t JT}[g,\Phi] = - S_0 \chi(\Sigma) - \bigg[ \frac{1}{2} \int_\Sigma \sqrt{g} \Phi \lr{R+2} + \underbrace{\int_{\pd \Sigma} \sqrt{h} \Phi \lr{K-1}}_{-I_{\t{bdy}}}\bigg],
\ee
where the first term is topological and $\chi(\Sigma) = 2-2g-b$ for surfaces with $g$ handles and $b$ boundaries. When we include asymptotic boundaries we choose boundary conditions to fix the induced metric $\sqrt{h}=1/\epsilon$ and the proper length of the boundary to be $\beta/\epsilon$, and we fix the dilaton to asymptotically approach $\phi_b = \gamma/\epsilon$.

To integrate over metrics we must first specify a measure on the moduli space of all metrics, that is we must specify a metric $\lb.\hspace{.06cm},.\rb$ on the tangent space of moduli space. The tangent space is naturally the space of metric deformations, and the standard metric for the space of these deformations is the ultra-local measure\cite{Polyakov:1981rd}
\be \label{ultralocalmeasure}
\lrb{\delta g_{a b}, \delta g_{c d}} = \mathcal{N} \int_\Sigma \sqrt{g} g^{a c} g^{b d} \delta g_{a b} \delta g_{c d},
\ee
which can be extended to other tensor deformations in an obvious way. The measure is defined up to a normalization constant $\mathcal{N}$ that we will fix later. Metric deformation decompose into three orthogonal parts
\be \label{eqn:metric_deformation}
\delta g_{ a b} = \underbrace{\omega g_{a b}}_{\t{weyl}} \hspace{.1cm}\oplus \hspace{.1cm} \underbrace{\operatorname{range} P_1}_{\t{small diff}} \hspace{.1cm} \oplus \hspace{.1cm} \underbrace{\operatorname{ker} P_1^{\dag} }_{\t{moduli}},
\ee
where $( P_1 V)_{a b} = \nabla_a V_b + \nabla_b V_a - (\nabla_c V^c ) g_{a b}$, and $(P_1^\dag \delta g)_a = -2 \nabla^b \delta g_{a b}$. A general infinitesimal metric deformation is a combination of a Weyl transformation, a small diffeomorphism, and a deformation of the moduli. Since the metric deformations are orthogonal \eqref{eqn:metric_deformation} the path integral measure breaks up into integrals over these deformations. 

We will first consider the case where we are integrating over all metrics on a compact surface of genus $g\geq2$. To perform the integral we gauge fix to conformal gauge and write the metric as $g=f^{*}\lr{e^{2\omega} \hat{g}}$ with $f$ a small diffeomorphism, $\omega$ a Weyl factor, and $\hat{g}$ a constant negative curvature metric $\hat{R}=-2$. The measure works out to be\cite{HokerPhong88,DHoker:1985een,Moore:1985ix}
\be \label{eqn:Z_intermediate1}
Z_g = \int_{\mathcal{M}_g} \underbrace{d\lr{\t{Weil-Pet.}} \hspace{.1cm}(\det \hat{P}_1^\dag \hat{P}_1)^{1/2} \mathcal{D} \omega}_{\mathcal{D} g/\t{Vol}} \mathcal{D} \Phi e^{-I_{\t JT}[e^{2\omega}\hat{g},\Phi]} e^{-26 S_L[\omega, \hat{g}]}.
\ee
The integral is carried out over the moduli space of constant negative curvature surfaces $\mathcal{M}_g$ of genus $g$.\footnote{When integrating over small diffeomorphisms we pick up a factor of the volume of small diffs. The ratio of small diffs with all diffs gives the set of large diffs, which reduces the integral $\frac{\text{Vol$($Diff$_0)$}}{\t{Vol(Diff)}}\int_{\mathcal{T}_g} = \int_{\mathcal{M}_g}$ to the moduli space of $R=-2$ surfaces. Here $\mathcal{T}_g$ is the set of metrics mod Weyl $[e^{2\omega} g]$ before quotienting by diffeomorphisms, also known as Teichmuller space.} In the above $(\det \hat{P}_1^\dag \hat{P}_1)^{1/2}$ appears from the ghost path integral when gauge fixing to conformal gauge, and the notation $\hat{P_1}$ implies the operator is defined with respect to $\hat{g}$. We have an integral over the Weyl factor $\mathcal{D} \omega$, and the Liouville action $S_L$ appears due to the conformal anomaly arising from defining the integration measures with respect to $\hat{g}$.\footnote{Every integral in \eqref{eqn:Z_intermediate1}, such as $\mathcal{D}_{\hat{g}} \Phi$, is defined with the appropriate ultra-local measure \eqref{ultralocalmeasure} with metric $\hat{g}$.} The Liouville action will give no contribution after we perform the $\mathcal{D} \omega$ integral so we discard it from now on. When restricted to constant negative curvature metrics $\hat{g}$, the measure \eqref{ultralocalmeasure} is by definition the Weil-Petersson measure \cite{hubbard2006teichmuller}, which is the origin of $d\lr{\t{Weil-Pet.}}$.

We can write the part of the JT gravity action \eqref{I_JT} that contains the dilaton as
\be
I[e^{2\omega}\hat{g},\Phi] = -\frac{1}{2}\int_\Sigma \sqrt{\hat{g}} \Phi \lr{\hat{R}-2\hat{\nabla}^2 \omega + 2 e^{2\omega}}.   
\ee
We can perform the integral over the dilaton by rotating the integration contour to be along the imaginary axis, giving a delta function constraint that localizes onto constant negative curvature geometries
\be
\int \mathcal{D} \omega \int_{i \mathbb{R}} \mathcal{D} \Phi e^{-I_{\t JT}[e^{2\omega}\hat{g},\Phi]} = \int \mathcal{D} \omega \hspace{.1cm}\delta \lr{\hat{R}-2\hat{\nabla}^2 \omega + 2 e^{2\omega} } = \frac{1}{\det(-2\hat{\nabla}^2 + 4)},
\ee
where in the last equality we localize onto $\omega=0$. A straightforward calculation shows that the gauge fixing differential operator takes the form $\hat{P}_1^\dag \hat{P}_1 = 2 \lr{-\hat{\nabla}_1^2+1}$ on a surface with constant negative curvature, with $\hat{\nabla}^2_1$ the vector laplacian. Putting everything together we find that the JT gravity path integral \eqref{eqn:Z_intermediate1} is
\begin{align}
Z_g &= (\text{const})^\chi \int_{\mathcal{M}_g} d\lr{\t{Weil-Pet.}} \frac{\det\lr{-\hat{\nabla}_1^2+1}^{1/2}}{\det(-\hat{\nabla}^2 + 2)}\\
&=(\text{const})^\chi \int_{\mathcal{M}_g} d\lr{\t{Weil-Pet.}}.
\end{align}
In the above we have discarded some multiplicative constants from the determinants, since in zeta function regularization multiplying all the eigenvalues by a constant shifts the determinant by $(\text{const})^\chi$.\footnote{All of these constants can be absorbed into a redefinition of $S_0$ in the topological term of the action \eqref{I_JT}.} Furthermore we also used our result for the vector laplacian determinant
\be
\frac{\det\lr{-\hat{\nabla}_1^2+1}^{1/2}}{\det(-\hat{\nabla}^2 + 2)} = 2^{\frac{1}{2}\chi},
\ee
given in equation \eqref{eqn:DetRatio_Final}. The above conclusion immediately generalizes beyond compact surfaces. 

Consider the moduli space of genus $g$ hyperbolic surface with $n$ asymptotic boundaries of regularized lengths $\vec{\beta} = \lr{\beta_i, \ldots, \beta_n}$ and $k$ conical defects with opening angles specified by $2\pi \alpha_i$. We take $\vec{\a} = \lr{\a_1, \ldots, \a_k}$ with $\alpha_i = n_i^{-1}$ and integer $n_i\geq2$. To localize onto geometries with such singularities we must insert an operator $\mathcal{V}_\a$ into the path integral \eqref{Z_JT}. We explain the dilaton gravity form of this operator slightly later, but for now in all integrals we assume we have inserted this operator and integrated over the dilaton to localize onto the relevant hyperbolic geometries. 

When integrating over hyperbolic surfaces with arbitrary conical singularities and asymptotic boundaries we again obtain the Weil-Petersson measure for the associated moduli spaces. This is because the measure \eqref{ultralocalmeasure} for metrics with conical singularities is again by definition the Weil-Petersson metric on such spaces\cite{schumacher2011weil}. Carrying out the JT gravity path integral in the same way as for a compact surface we arrive at
\begin{align}
Z_{g,{\Vec{\alpha}}} (\beta_1,\ldots, \beta_n) &= (\text{const})^\chi \int_{\mathcal{M}_{g,{\Vec{\alpha},\Vec{\beta}}}} d\lr{\t{Weil-Pet.}} \frac{\det\lr{-\hat{\nabla}_1^2+1}^{1/2}}{\det(-\hat{\nabla}^2 + 2)} e^{-I_{\t{bdy}}} \nn \\
&=(\text{const})^\chi\int_{\mathcal{M}_{g,{\Vec{\alpha},\Vec{\beta}}}} d (\text{boundary wiggles}) d\lr{\t{bulk moduli}} e^{-I_{\t{bdy}}}.
\end{align}
In the above we have split the integral over the moduli $d\lr{\t{Weil-Pet.}}$ into the boundary wiggles and the bulk moduli. The Weil-Petersson measure \eqref{ultralocalmeasure} will induce a measure for the boundary wiggles, which we will calculate in section \ref{sec:disk} and see is given by the standard schwarzian measure. While we are only able to compute the required determinants for special defect angles, it seems likely that the cancellation between determinants will remain the case for general conical defects.

\subsubsection{Integrating over moduli space} \label{intmodulispace}
We now explain how to formally carry out the integration over moduli space. For reviews see \cite{DHoker:1985een,HokerPhong88,Giddings:tasi,GidHok87,Polchinski:1998rq}. From \eqref{eqn:metric_deformation} the infinitesimal deformations of the moduli correspond to variations of the metric $\delta g \in \operatorname{Ker} P_1^\dag$. In conformal gauge where the metric is off-diagonal this translates to the condition $\ol \pd \delta g_{z z} = \pd \delta g_{\bar z \bar z}=0.$ This implies that the moduli deformations are holomorphic/anti-holomorphic deformations of the form
\be
\phi_n = \phi_n(z) dz^2,\qquad \ol{\phi_n} = \ol{\phi_n} (\bar z) d\bar z^2 . 
\ee 
These are known as quadratic differentials, and they provide a basis for infinitesimal deformations of the moduli.

We briefly explain some properties of quadratic differentials. On a compact genus $g\geq2$ surface there is a basis of $3g-3$ globally defined pairs of holomorphic and anti-holomorphic quadratic differentials giving a moduli space of real dimension $6g-6$. When we consider hyperbolic surfaces with conical defects, the quadratic differentials are allowed to have simple poles at the location of the defect $\phi \sim z^{-1} dz^2$ \cite{schumacher2011weil}.\footnote{For examples of quadratic differential on surfaces with defects see \cite{eynard2018lectures}.}
These deformations correspond to moving the conical defect on the surface, and so the real dimension of the moduli space with $n$ conical defects is $6g-6+2n$. When we consider the inclusion of asymptotic boundaries the boundary fluctuations are also part of the moduli, and there are infinitely many quadratic differentials associated to turning on modes of the boundary fluctuations as we will see in section \ref{sec:disk}. 

An arbitrary infinitesimal moduli deformation, in conformal gauge, can be written in terms of quadratic differentials as\footnote{The normalization factor of \eqref{eqn:quadratic_diff_inner} is our convention. We will fix the normalization factor of the metric perturbation inner product later, and are satisfied with the proportionality here as a motivation.}
\be \label{eqn:deltag_intro3}
\delta g = \sum_n \delta c_n \phi_n(z) dz^2 + \delta \ol c_n \ol \phi_n(\bar z) d \bar{z}^2,
\ee
where $\delta c_n$ are deformation parameters. Using the inner product for metric deformations \eqref{ultralocalmeasure} we can define an inner product for quadratic differentials, given by 
\be\label{eqn:quadratic_diff_inner}
\lrb{\phi_n, \phi_m} \equiv \int_\Sigma d^2 z \sqrt{g} g^{z \ol z} g^{z \ol z} \phi_n(z) \ol{\phi}_m(\bar z),
\ee
where the second quantity is to be complex conjugated. It is standard to ignore the normalization in \eqref{ultralocalmeasure}, restoring it only when computing the full path integral measure. 

To define the measure on moduli space we will need one more quantity known as the Beltrami differential $\mu$. To integrate over moduli space we must first choose coordinates $m_n, \bar{m}_n$ on the space  by specifying a set of metrics $g(z, \bar{z} ;m_n, \bar{m}_n)$ that give a slice through the space. We must then compute the Jacobian for this set of coordinates. This can be accomplished as follows. Consider a metric $g(m_n, \bar{m}_n)$ on a surface which in some local patch takes form $ds^2 = e^{2\omega} |dz|^2=2 g_{z\bar{z}}|dz|^2$ for some complex coordinate $z$. As we change the moduli by moving to a nearby metric at $m_n + \delta m_n$ the new metric can no longer be $\propto |dz|^2$ since it is not in the same equivalence class as the original metric. However, there must be complex coordinates $z'=z+\delta m_n \, v(z,\bar{z}) + \ldots$ where the new metric takes the form $e^{2(\omega + \delta \omega)}|dz'|^2$. We thus find that the deformed metric in the old $z$ coordinates is
\be
ds'^2 = e^{2 (\omega + \delta \omega)} \lr{dz d \bar{z} + \delta m_n \hspace{.07cm} \ol{\mu}_n dz^2 + \delta \ol{m}_n \hspace{.07cm} \mu_n d\bar{z}^2 + \ldots}, \qquad
\delta (ds^2) = \delta m_n e^{2\omega} \ol \mu_n dz^2 + \text{c.c.} 
\ee
where we have defined the Beltrami differentials $\mu = \ol \pd v$, and $\ol{\mu} = \pd v$,\footnote{The Beltrami differential are commonly viewed as $(-1,1)$ forms $\mu = \mu_{\bar z}^{~z} d z^{-1} d \bar z$ which can be integrated against $(2,0)$ forms $\phi_{zz} dz^2$, and it is convention to define the metric deformation by beltrami differentials as $\ol \mu dz^2$.} which capture how the metric infinitesimally changes as we change the $m_n$ coordinates. In the second equation we have written the variation to linear order.\footnote{All these statements are up to Weyl rescalings since rescalings do not move the metric in moduli space.} The overlap between the quadratic differentials and the change in the metric due to a deformation in $m_n$ coordinates can be obtained from the measure \eqref{ultralocalmeasure}, which is most commonly expressed as an overlap between the Beltrami differentials $\mu$ and the quadratic differentials
\be\label{eqn:muphiprod}
\lrb{\mu_n, \phi_m} \equiv 2 \int_\Sigma d^2 z \mu_n \hspace{.05cm}\phi_m,
\ee
where the overlap is defined without a conjugation. Notice that deforming the metric by a quadratic differential $\phi(z) dz^2$ \eqref{eqn:deltag_intro3} is equivalent to deforming it by a beltrami differential $\ol \mu = \frac{\phi}{2 g_{z \bar z}}$ to linear order in the deformation. To compute the correct measure we must project the above deformation onto the space of genuine moduli deformations (as opposed to diffeomorphisms and Weyl rescalings). Projecting onto the basis of quadratic differentials we get
\be
\delta g = \delta m_n \phi_j \lrb{\ol \mu_n, \ol \phi_i}\lrb{\phi_j,\phi_i}^{-1}  + \delta \ol m_n \ol{\phi}_j \lrb{\mu_n,\phi_i} \lrb{\ol\phi_j,\ol \phi_i}^{-1},
\ee
where the notation $\lrb{...}^{-1}$ is an inverse matrix. Using \eqref{ultralocalmeasure} we can compute the metric
\be
ds^2_{\t{Weil-Pet.}} = 2 \mathcal{N} \lrb{\ol \mu_n , \ol\phi_j} \lrb{\mu_m , \phi_i} \lrb{\phi_i, \phi_j}^{-1} d m_n d \ol m_m.
\ee
Taking the square root of the determinant and treating the metric as a product of matrices, we find that in terms of $m_n$ coordinates the integral over moduli space takes the form
\be \label{WP_cncoords}
\int_{\mathcal{M}} d (\t{Weil-Pet.}) \hspace{.1cm} e^{-I} = \prod_n \mathcal{N} \int d m_n d \ol m_n  \frac{\det \lrb{\mu ,\phi} \det \lrb{\ol{\mu},\ol \phi} }{\sqrt{\det \lrb{\phi, \phi} \det \lrb{\ol \phi, \ol \phi}}} \hspace{.1cm} e^{-I[m_n, \ol m_n]}.
\ee
We have also included a possibility of some action $I$ that depends on the moduli. In JT gravity this action will be given by the boundary term in \eqref{I_JT}. We have also reinstated the overall normalization constant $\mathcal{N}$ appearing in the measure \eqref{ultralocalmeasure}. In \cite{Moitra:2021uiv}, and in appendix \ref{app:DTmeasure} using quadratic differentials, it can be shown that in the second order formalism the gluing measure for geodesic boundaries is given by
\be
4 \mathcal{N} \int b d b,
\ee
and so to get standard form of the Weil-Petersson measure we must choose the normalization $\mathcal{N}=\frac{1}{4}$. We keep the normalization constant present throughout, inserting it's numerical value when necessary.

\subsubsection*{Conical defect operator}
From the expression for the Weil-Petersson measure we can extract the dilaton gravity operator that creates a conical defect on the surface. When we have a defect we have two additional moduli that move the defect. We choose the moduli to be parameterized by the location $z_i$ of the defect, and we can formally write the measure as
\be \label{eqn:conicaldefectmeasure}
\mathcal{N} \int_\Sigma d^2 z_i \frac{\lb \mu ,\phi_1 \rb \lb \ol \mu , \ol \phi_1 \rb}{\sqrt{\lb \phi_1, \phi_1 \rb \lb \ol \phi_1, \ol \phi_1 \rb}},
\ee
where $\mu$ is the beltrami differential that infinitesimally moves the defect to $z_i + \delta z_i$, and $\phi_1$ is the quadratic differential with a simple pole at $z_i$. From the JT gravity perspective this is the measure after the dilaton has already been integrated out, and so restoring it we arrive at the dilaton gravity operator that creates a conical defect
\be
\mathcal{V}_\a = \mathcal{N} \int_\Sigma d^2 z_i \frac{\lb \mu ,\phi_1 \rb \lb \ol \mu , \ol \phi_1 \rb}{\sqrt{\lb \phi_1, \phi_1 \rb \lb \ol \phi_1, \ol \phi_1 \rb}} e^{-2\pi (1-\a)\Phi(z_i)}.
\ee
One complication for evaluating the above integral is that quadratic differentials are not generally known, apart from their singular behavior near the defect. However, for the hyperbolic disk everything can be worked out explicitly, and in section \ref{sec:blunt_disk_z0} we evaluate the measure on the disk. In section \ref{sec:blunt_generalsurfaces} we argue that for general surfaces in the blunt defect limit $\a \to 1$ the measure can be worked out to leading order in the $(1-\a)$ expansion by using the universal divergent behavior of the quadratic differential.

\subsection{Examples} \label{sec:3.2examples}
\subsubsection{Disk} \label{sec:disk}
In this section we explain how to apply the above formalism for integrating over moduli space to reproduce disk partition function $Z(\beta)$ in JT gravity\cite{Charles:2019tiu,Choi:2021nnq,Choi:2023syx}. The AdS$_2$ disk metric in complex coordinates is given by\footnote{Our conventions for complex coordinates are $z=x+i y$ with $\int d^2 z = 2 \int d^2 x$. The dirac delta function is defined by $\ol \pd \pd \log |z|^2 = 2\pi \delta^2 (z, \bar z)$, with $\int d^2 z \delta^2 (z, \bar z) = 1$. A Weyl rescaled metric $g = e^{2\omega} d z d \bar{z}$ has curvature $R = -8 e^{-2\omega}\ol \pd \pd \omega$.}
\be \label{eqn:gAdS2_zcoords}
ds^2 = \frac{4 \dz}{ \lr{1-|z|^2}^2}.
\ee
This is related to the standard metric on the hyperbolic disk through the coordinate change
\be
ds^2 = d r^2 + \sinh^2 r \hspace{.06cm} d\tau^2, \qquad z = e^{i \tau} \tanh \frac{r}{2}. 
\ee
As explained previously, deformations of the moduli correspond to holomorphic quadratic differentials and their conjugates. On the disk it is simple to write down a basis of holomorphic functions given by
\be \label{eqn:QHD_Disk}
\phi_n = \sqrt{\frac{n^3-n}{2\pi}}z^{n-2} dz^2, \qquad \ol \phi_n = \sqrt{\frac{n^3-n}{2\pi}} \ol z^{n-2} d\ol z^2, \qquad \lb \phi_n, \phi_m \rb = \delta_{n,m},
\ee
where for holomorphicity we restrict to integers $n \geq 2$. Infinitesimal deformations of the disk are given by deforming by linear combinations of the quadratic differentials
\be \label{eqn:g_QHD_deformation}
ds^2 = \frac{4 \dz}{ \lr{1-|z|^2}^2} + c_n \sqrt{\frac{n^3-n}{2\pi}}z^{n-2} dz^2 + \ol c_n \sqrt{\frac{n^3-n}{2\pi}} \ol z^{n-2} d\ol z^2.
\ee
Note that the above metric is no longer AdS$_2$. We will take our coordinates for moduli space to be given by $c_n, \ol c_n$, and so we must work out the measure for these coordinates along with the boundary action of JT gravity \eqref{I_JT} in terms of $c_n,\ol c_n$. The measure is straightforward since for the Beltrami differentials in \eqref{WP_cncoords} we just use the dual of the quadratic differentials $\ol \mu_n = \frac{\phi_n}{2 g_{z \ol z}}$. Working out the action is slightly more complicated, which we now do.

The finite version of the metric for the above deformation was worked out by \cite{Choi:2023syx}, and is given by\footnote{We mention that we still mostly work with infinitesimal perturbations of the metric and the following finite metric deformation is not unique. }
\begin{align} \label{eqn:g_QHD_finite}
& ds^2  = e^{C + \ol C}\frac{\lr{1-|z|^2}^2}{\lr{1-|e^C z|^2}^2}\left[\lr{|1+z \pd C|^2 + | \ol z \ol \pd C|^2 } \frac{4 \dz}{\lr{1-|z|^2}^2} {+} \lr{1+ z \pd C} \ve'''(z) dz^2 + \t{c.c.}  \right],\\ & C(z,\ol z) = {{-}}\frac{1}{2 z}\left(\ve(z)-z^2 \ol{\ve}(\ol{z})-z(1-z \ol{z})\ol{\ve}'(\ol{z})-\frac{1}{2}(1-z \ol{z})^2 \ol{\ve}''(\ol{z})\right). \nn
\end{align}
The functions $\ve(z), \ol \ve (\ol z)$ are holomorphic/anti-holomorphic, and the primes $\ve'(z), \ol \ve'(\ol z)$ denote holomorphic/anti-holomorphic derivatives. The metric \eqref{eqn:g_QHD_finite} is AdS$_2$, which can be seen by using coordinates $w=z e^C$ in terms of which it takes the standard form \eqref{eqn:gAdS2_zcoords}. This coordinate change is a large diffeomorphism and corresponds to a physically distinct configuration. The relation between \eqref{eqn:g_QHD_deformation} and \eqref{eqn:g_QHD_finite} comes from identifying 
\be\label{eqn:epsilon_mode}
\ve(z) = \sum_{n \geq 2} \frac{1}{\sqrt{2\pi(n^3 - n)}} \, c_n z^{n+1}, \qquad \ol\ve(\ol z) = \sum_{n \geq 2} \frac{1}{\sqrt{2\pi (n^3-n)}} \,\ol c_n \ol z^{n+1}.  
\ee
Expanding the metric \eqref{eqn:g_QHD_finite} to leading order in $\ve$ we recover the infinitesimal form \eqref{eqn:g_QHD_deformation} up to a Weyl factor of order $\varepsilon^2$.

We now explain how turning on these deformations corresponds to turning on modes of the Schwarzian. In $z$ coordinates \eqref{eqn:g_QHD_finite} the boundary cutoff is located at fixed radial distance while in $w=z e^C$ coordinates the metric takes the form 
\be
ds^2 = \frac{4 d w d \ol w}{(1-|w|^2)^2},
\ee
and the cutoff surface fluctuates. Taking $z=e^{i \theta}$ we find that at the boundary and linear order of $c_n$
\be \label{eqn:w(theta)}
w = e^{i f(\theta)}, \qquad
f(\theta) = \theta + \sum_{n \geq 2} \frac{1}{2i \sqrt{2\pi (n^3-n)}} \lr{c_n e^{i n \theta}-\ol c_n e^{-i n \theta}}. 
\ee
We would like to find the cutoff surface in terms of the $f(\theta)$ reparameterization. We can do this perturbatively in the cutoff $\delta$ finding $w\lr{\theta} = e^{i f(\theta)} \lr{1 - \delta f'(\theta) + \frac{\delta^2 f'(\theta)^2}{2} + \mathcal{O}(\delta^3)}$. Where we have found the radial distance in terms of $f(\theta)$ using that the induced metric along the cutoff surface is fixed to be $\sqrt{h}= 1/\delta$. The extrinsic curvature can then be evaluated 
\begin{align}
K= 1+\delta^2 \lr{\frac{f'''}{f'} - \frac{3 f''^2}{2 f'^2 }+ \frac{1}{2} f'^2} + \mathcal{O}(\delta^3) = 1+\delta^2 \operatorname{Sch}\lr{\tan \frac{f(\theta)}{2}, \theta},
\end{align}
which is the standard Schwarzian answer. From \eqref{eqn:w(theta)} we see that the quadratic differentials $\phi_n, \ol \phi_n$ precisely correspond to turning on the modes of the Schwarzian. The boundary action \eqref{I_JT} can now be evaluated. Reinstating the temperature $\beta$ and the value of the dilaton $\phi = \gamma/\delta$ we find
\begin{align}
I_{\t{bdy}} &=-\frac{2 \pi \gamma}{\beta} \int_0^{2\pi} d \theta \lr{\frac{f'''}{f'} - \frac{3 f''^2}{2 f'^2 }+ \frac{1}{2} f'^2},\\
&=I_{0} + \frac{\pi \gamma}{2\beta} \sum_{n \geq 2} n \hspace{.01cm}c_n \ol c_n, \qquad  I_{0}=-\frac{2\pi^2\gamma}{\beta}. \nn
\end{align}
Where in the second line we have used that the Schwarzian path-integral is one-loop exact \cite{Stanford:2017thb} so that it is sufficient to expand to quadratic order in $c_n, \ol c_n$ and carry out the integral\cite{Saad:2019lba,Sarosi:2017ykf}, and $I_0$ is the classical contribution. We can now evaluate the path integral over the moduli space using \eqref{WP_cncoords}
\begin{align} \label{eqn:disk_calculation}
Z_{\t{disk}}\lr{\beta} &= \prod_{n \geq 2}\mathcal{N} \int d c_n d \ol c_n \left| \frac{\det \lb \mu ,\phi \rb}{\sqrt{\det \lb \phi, \phi \rb}} \right|^2 e^{-I_{\t{bdy}}} 
= e^{\frac{2\pi^2 \gamma}{\beta}} \prod_{n \geq 2} \mathcal{N} \int d c_n d \ol c_n \exp\lr{-\frac{\pi \gamma}{2 \beta} \sum_{n\geq2} n \hspace{.01cm} c_n \ol c_n}\\
&= e^{\frac{2\pi^2 \gamma}{\beta}} \prod_{n\geq2} \mathcal{N} \frac{ 4 \beta}{\gamma n} = \frac{1}{8}\sqrt{\frac{\gamma^3}{2\pi \mathcal{N}^3 \beta^3}}
\hspace{.1cm}e^{\frac{2\pi^2 \gamma}{\beta}}. \nn
\end{align}
In the first line we used our normalization for the quadratic differentials $\lb \phi_n, \phi_m \rb=\delta_{n,m}$ and that the dual beltrami differentials $\ol \mu_n = \frac{\phi_n}{2 g_{z \ol z}}$ satisfy $\lb \mu_n, \phi_m \rb = \delta_{n,m}$. In the last line we used zeta function regularization. This reproduces the JT gravity disk partition function obtained from the first order formalism \cite{Saad:2019lba}. 

\subsubsection{Conical defect}
We now explain how the above calculation is modified in the case of a disk with a single conical defect of opening angle $2\pi \a$. If the defect is at the center of the disk then the metric and Ricci scalar are given by
\be \label{eqn:ds2defect}
ds^2 = \frac{4 \a^2 |z|^{2(\a-1)}}{(1-|z|^{2 \a})^2} dz d\bar{z}\,,\qquad \frac{1}{2} \sqrt{g} \lr{R+2} = 2\pi(1-\alpha) \delta^{(2)}(z). 
\ee
The quadratic differentials are now given by
\be \label{eqn:defect_QHD_definition}
\phi_n = \sqrt{\frac{n^3- n\alpha^2}{2\pi}}\hspace{.1cm} z^{n-2}, \qquad \ol \phi_n = \sqrt{\frac{n^3-n \alpha^2}{2\pi}} \hspace{.1cm} \ol z^{n-2} d \ol z^2,\qquad \lb\phi_n, \phi_m \rb = \delta_{n,m}.
\ee
When a conical defect is present on the geometry the quadratic differentials are allowed to have simple poles\cite{schumacher2011weil} at the location of the defect so we now allow for $n \geq 1$ modes. This increases the real dimension of the moduli space by two, and the new deformation should be interpreted as moving the defect on the surface. The quadratic differentials infinitesimally deform the metric by
\be \label{eqn:g_QHD_deformation_defect}
ds^2 = \frac{4 \a^2 |z|^{2(\a-1)}}{(1-|z|^{2 \a})^2} dz d\bar{z} + c_n  \sqrt{\frac{n^3-n\alpha^2}{2\pi}}\hspace{.1cm} z^{n-2} dz^2 + \ol c_n \sqrt{\frac{n^3-n \alpha^2}{2\pi}} \hspace{.1cm} \ol z^{n-2} d\ol z^2.
\ee
Similar to the case of the disk, there is a vector field $\xi$ that infinitesimally removes these deformations, and we can find it by solving for
\be
2 \nabla_z \xi_z = -c_n \sqrt{\frac{n^3-n\alpha^2}{2\pi}}\hspace{.1cm} z^{n-2}, \qquad 2 \nabla_{\bar z} \xi_{\bar z} = -\ol c_n \sqrt{\frac{n^3-n\alpha^2}{2\pi}}\hspace{.1cm} \ol z^{n-2}, \qquad \nabla_z \xi_{\bar{z}} + \nabla_{\bar z} \xi_{z}=0.
\ee
The vector field can be found and takes the form
\begin{align}\label{eqn:defectvectorfield}
\xi^{\bar z} = \frac{\ol c_n \ol z^{n+1}}{2\sqrt{2\pi(n^3-n\a^2)}}-\frac{c_n \ol z (z \ol z )^{-\a} \left(2 \a^2 z^n (z \ol z)^\a+n^2 z^n \left(1-(z\ol z )^\a \right)^2+ n\a z^n \left(1-(z \ol z )^{2 \a}\right)\right)}{4 \a^2 \sqrt{2\pi(n^3-n\a^2})},\\
\xi^{ z} = \frac{c_n  z^{n+1}}{2 \sqrt{2\pi(n^3-n\a^2)}}-\frac{c_n  z (z \ol z )^{-\a} \left(2 \a^2 \ol z^n (z \ol z)^\a+n^2 \ol z^n \left(1-(z\ol z )^\a \right)^2+ n\a \ol z^n \left(1-(z \ol z )^{2 \a}\right)\right)}{4 \a^2 \sqrt{2\pi(n^3-n\a^2})}. \nn
\end{align}
The finite form of the deformed metric can be recovered by choosing coordinates $w=z \exp\lr{\frac{1}{z}\xi^z}$ where in $w$ coordinates we have the metric \eqref{eqn:ds2defect} (compare to \eqref{eqn:g_QHD_finite}), and we implicitly sum over $c_n, \ol c_n$ in $\xi^z$. We can now follow the same procedure to find the shape of the cutoff in $w$ coordinates as we followed for the disk. Following the curve $z=e^{i \theta}$ we find in $w$ coordinates
\be
w=e^{i f(\theta)}, \qquad f(\theta) = \theta + \sum_{n \geq 1} \frac{1}{2i \sqrt{2\pi (n^3-n \a^2)}} \lr{c_n e^{i n \theta}-\ol c_n e^{-i n \theta}}.
\ee
Fixing the induced metric along the cutoff surface, we again find the same functional form with a defect present $w\lr{\theta} = e^{i f(\theta)} \lr{1 - \delta f'(\theta) + \frac{\delta^2 f'(\theta)^2}{2} + \mathcal{O}(\delta^3)}$. Calculating the extrinsic curvature along this curve we find
\begin{align}
K= 1+\delta^2 \lr{\frac{f'''}{f'} - \frac{3 f''^2}{2 f'^2 }+ \frac{\a^2}{2} f'^2} + \mathcal{O}(\delta^3) = 1+\delta^2 \operatorname{Sch}\lr{\tan  \frac{\a f(\theta)}{2}, \theta}.
\end{align}
Which is precisely the Schwarzian with a conical defect\cite{Mertens:2019tcm} which is again one-loop exact. Restoring temperature and the dilaton as in the case of the disk we find the action to quadratic order to be
\begin{align} \label{eqn:Ibdy_defect}
I_{\t{bdy}} &=-\frac{2 \pi \gamma}{\beta} \int_0^{2\pi} d \theta \lr{\frac{f'''}{f'} - \frac{3 f''^2}{2 f'^2 }+ \frac{\a^2}{2} f'^2},\\
&=I_{0} + \frac{\pi \gamma}{2\beta} \sum_{n \geq 1} n \hspace{.01cm}c_n \ol c_n, \qquad  I_{0}=-\frac{2\pi^2\gamma}{\beta} \a^2 . \nn
\end{align}
Integrating over the moduli space with \eqref{WP_cncoords} we need to now include the $n=1$ mode compared with the disk calculation
\begin{align} \label{Z_defect_cn}
Z_{\t{defect}} \lr{\beta, \alpha} &= \prod_{n \geq 1} \mathcal{N} \int d c_n d \ol c_n \left| \frac{\det \lb \mu ,\phi \rb}{\sqrt{\det \lb \phi, \phi \rb}} \right|^2 e^{-I_{\t bdy}}=e^{\frac{2\pi^2 \gamma}{\beta}\alpha^2} \prod_{n \geq 1} \mathcal{N} \int d c_n d \ol c_n \exp\lr{-\frac{\pi \gamma}{2\beta} \sum_{n\geq1} n \hspace{.01cm} c_n \ol c_n} \\
&=e^{\frac{2\pi^2 \gamma}{\beta}\alpha^2} \prod_{n \geq 1} \mathcal{N} \frac{4\beta}{\gamma n} = \sqrt{\frac{\gamma}{8\pi \mathcal{N} \beta}} e^{\frac{2\pi^2 \gamma}{\beta} \alpha^2}, \nn
\end{align}
where again the measure simplifies due to our normalization \eqref{eqn:defect_QHD_definition} as in the case of the disk \eqref{eqn:disk_calculation}. This again matches the calculation of the partition function with a single defect \cite{Mertens:2019tcm,Saad:2019lba}. 

\subsubsection{Conical defect: $z$ coordinates} \label{sec:blunt_disk_z0}
In the previous section we integrated over moduli space using $c_n$ coordinates which correspond to deformations of the metric by quadratic differentials. In this section we will carry out the previous calculation using a coordinate choice that is more physically intuitive, namely the position $z_i$ of the defect on the disk. This choice of coordinates is applicable to the $n=1$ mode of the Schwarzian which can be thought of as moving the defect on the disk. To integrate over the moduli space with $z_i$ coordinates we must again work out both the measure and the action in \eqref{WP_cncoords}, which we now do. 

The metric with a conical defect at a general position $z_i$ can be obtained by applying an $\text{SL}(2,\mathbb{R})$ transformation to the metric with a defect at the center
\be \label{eqn:ds2_defect_z0}
ds^2 = \frac{4 \a^2 |w(z)|^{2(\a-1)}}{(1-|w(z)|^{2 \a})^2} |w'(z)|^2 dz d\bar{z}\,,\qquad w(z) = \frac{z-z_i}{1-z \ol z_i},
\ee
where in $w$ coordinates the defect is at $w=0$, but is at $z=z_i$ in $z$ coordinates. To work out the measure in \eqref{WP_cncoords} we need to know the Beltrami differential $\mu$ which contains information on how the metric infinitesimally changes as we move the defect $z_i \to z_i + \delta z_i$. As explained in section \ref{intmodulispace}, this can be obtained by finding a set of coordinates $z'(z,\bar{z})$ where the deformed metric is flat $\propto |dz'|^2$ and the defect is mapped to the new point $z'(z_i)=z_i + \delta z_i$ with the asymptotic boundary undeformed.\footnote{The coordinates $z'$ should not affect the boundary of the disk since such a deformation would both move the defect and turn on some additional Schwarzian modes, and we want to isolate the deformation that moves the defect.} To extract the Beltrami we only need this coordinate chart to linear order in $\delta z_i$, where it is given by
\be
z'(z,\bar{z})=\frac{\delta z_i \frac{1-z \bar{z}}{1-z_i\bar{z_i}}+z(1-z_i \bar{z})}{\delta z_i \bar{z}_i \frac{1-z \bar{z}}{1-z_i\bar{z_i}}+(1-z_i \bar{z})}.
\ee
\be
|dz'|^2 \propto |dz + \delta z_i \hspace{.04cm} \mu d \bar z + \mathcal{O}((\delta z_i)^2)|^2, \qquad \mu = -\frac{(z-z_i)(1-z \bar{z}_i)}{(1-z_i\bar{z}_i)(1-\bar{z} z_i)^2}\,.
\ee
Additionally, the quadratic differential that moves the defect when it is located $z=z_i$ is given by\footnote{It can be checked that $\lrb{\mu, \phi_n}=0$ for $n\geq2$, implying that the only action of $\mu$ is to move the defect.}
\begin{equation} \label{eqn:phi1_disk_defect}
    \phi_{1}=\frac{1}{w} (\pd w)^2 dz^2=\frac{(1-z_i \bar{z}_i)^2}{(z-z_i)(1-z \bar{z}_i)^3}dz^2, \qquad \lrb{\phi_{1}, \phi_{1}}= \frac{2\pi}{1-\a^2}.
\end{equation}
Which is precisely $\phi_1=\frac{1}{w} dw^2$ in $w$ coordinates, implying the norm relation noted above. Using the Beltrami and quadratic differential we can compute the overlap
\begin{equation} \label{eqn:diskmuphi}
    \langle \mu,\phi_1\rangle=2\int dz d\bar{z} \frac{1-|z_i|^2}{(1-z\bar{z}_i)^2(1-z_i\bar{z})^2}= \frac{4\pi}{1-|z_i|^2}\,.
\end{equation}
We can now immediately compute the Weil-Petersson measure \eqref{WP_cncoords} for the $z_i$ coordinates that move the defect
\begin{equation}\label{eqn:defectpositionmeasure2}
    \mathcal{N} \int d^2 z_i \left| \frac{\lb \mu ,\phi_1 \rb}{\sqrt{\lb \phi_1, \phi_1 \rb}} \right|^2 = 8 \pi \mathcal{N} \int d^2 z_i \frac{1-\a^2}{(1-|z_i|^2)^2}= 4 \pi \mathcal{N}  (1-\a^2) \int d^2 z_i \sqrt{g}  \,,
\end{equation}
where in the last equality we have noticed the measure takes the form of the metric on the disk without a defect. We are now left to compute the action in the $z_i$ coordinates. We follow the same procedure as in the previous section, if we take the cutoff to be at $z=e^{i \theta}$ then in $w=e^{i f(\theta)}$ coordinates \eqref{eqn:ds2_defect_z0} we have 
\begin{equation}
    w(e^{i\theta})=e^{i\theta}\frac{1-z_ie^{-i\theta}}{1-\bar{z}_i e^{i\theta}}\  \Rightarrow \ f(\theta)=\theta+ i^{-1} \log\left(\frac{1-z_ie^{-i\theta}}{1-\bar{z}_i e^{i\theta}} \right)\,.
\end{equation}
Note that this reparameterization only takes into account the $n=1$ mode of the Schwarzian. We previously worked out the action for $f(\theta)$ in \eqref{eqn:Ibdy_defect}, giving
\begin{align} \label{eqn:I_defect_z0}
I^{n=1}_{\t{bdy}} &=-\frac{2 \pi \gamma}{\beta} \int_0^{2\pi} d \theta \lr{\frac{f'''}{f'} - \frac{3 f''^2}{2 f'^2 }+ \frac{\a^2}{2} f'^2}, \nn\\
&=-\frac{2\pi\gamma}{\beta}\int_0^{2\pi} d\theta \frac{1}{2}\left[ \alpha^2 \frac{(1-|z_i|^2)^2}{|1-z_i e^{-i\theta}|^4}-\frac{(z_ie^{-i\theta}+\bar{z}_i e^{i\theta}-2)(z_ie^{-i\theta}+\bar{z}_i e^{i\theta}-2 |z_i|^2)}{|1-z_i e^{-i\theta}|^4}\right], \nn \\
&= -\frac{2\pi^2 \gamma}{\beta} \lr{1-(1-\a^2) \frac{1+|z_i|^2}{1-|z_i|^2} }.
\end{align}
We can now write the full answer using \eqref{WP_cncoords}, \eqref{eqn:defectpositionmeasure2}, and \eqref{eqn:I_defect_z0}. Including all the $n \geq 2$ modes of the Schwarzian we get
\be
Z_{\t{defect}}(\beta, \alpha) =4\pi \mathcal{N} \int d^2 z_i \frac{\lr{1-\a^2} }{\lr{1-|z_i|^2}^2} e^{-I^{n=1}_{\t{bdy}}} \times \prod_{n\geq 2} \mathcal{N} \frac{4 \beta}{\gamma n} = e^{\frac{2\pi^2 \gamma}{\beta} \alpha^2} \prod_{n\geq 1} \mathcal{N} \frac{4\beta}{\gamma n},
\ee
which is the expected answer \eqref{Z_defect_cn}. From the above we see that when integrating over the position of the defect we pick up a measure factor proportional to $(1-\a^2)$ along with an action proportional to the same. In the limit $\a \to 1$ the defect becomes very blunt and one expects that it does not backreact strongly on the geometry, so that semiclassical methods can be applied. However, we see that the path integral measure is also becoming important in this limit, and must be taken into account. 

We can rewrite the integral over $z_i$ in a way that connects it to the semiclassical calculation of a single defect on the disk. The contribution of the $n=1$ mode of the Schwarzian is
\be \label{eqn:n=1Schwarzian}
Z^{n=1}_{\t{defect}}(\beta, \alpha) = 4\pi \mathcal{N} (1-\a^2) \int d^2 z_i \sqrt{g(z_i)} e^{-\pi (1-\alpha^2) \phi_{\t{cl.}}(z_i)}, \qquad \phi_{\t{cl.}}(z_i)=\frac{2\pi \gamma}{\beta}\frac{1+|z_i|^2}{1-|z_i|^2},
\ee
where in the above $\sqrt{g}$ is the metric on the hyperbolic disk without a defect and $\phi_{\t{cl.}}$ is the classical solution for the dilaton on the disk. It has previously been observed that a semiclassical calculation \cite{Witten:2020wvy,Maxfield:2020ale} of the disk with a single defect seems to agree with the full Schwarzian calculation if an appropriate measure factor is inserted for the integral over the dilaton. The semiclassical calculation is given by
\be
\lim_{\a \to 1} Z_{\t{defect}}^{\t{semiclassical}} \approx 2\pi \lr{1-\a} \int d^2 z \sqrt{g} e^{-2\pi \lr{1-\a} \phi_{\t{cl.}}(z)},
\ee
where we approximate the integral over the dilaton by it's classical value, and a measure factor of $2\pi\lr{1-\a}$ is inserted by hand to get agreement with the Schwarzian. Comparing with \eqref{eqn:n=1Schwarzian} we see why this semi-classical calculation works, the integral over the position cancels out the measure in both calculations giving the same answer.

\subsubsection{General surfaces} \label{sec:blunt_generalsurfaces}
In this section we will give a general argument that the correct form of the path integral measure for a conical defect in the limit that the angle becomes blunt is given by
\be 
\lim_{\a \to 1} \mathcal{N} \int d^2 z \left| \frac{\lb \mu ,\phi_1 \rb}{\sqrt{\lb \phi_1, \phi_1 \rb}} \right|^2 = 2\pi (1-\a)\int d^2 z \sqrt{g(z)} + \mathcal{O}\lr{\lr{1-\alpha}^2},
\ee
where we have inserted $\mathcal{N}=\frac{1}{4}$ to normalize the operator with respect to the standard gluing measure $\int b db$. 
Consider a surface $\Sigma$ with at least one conical defect\footnote{For simplicity, we do not consider geodesic boundary temporarily.} located at $z=z_i$. The quadratic differential that moves this defect has a simple pole at $z_i$
\begin{equation}
    \phi_{1}=\frac{1}{z-z_i} dz^2 + \ldots\,,
\end{equation}
where the subleading terms are holomorphic on $\Sigma$.\footnote{We can also include a residue at the pole, as we did for the disk in \eqref{eqn:phi1_disk_defect} and \eqref{eqn:diskmuphi}, but it will cancel out of the path integral measure.} In principle we need to know the quadratic differential on the entire surface to calculate the measure, but in the blunt defect limit $\a \to 1$ we will argue that the singular behavior of the quadratic differential dominates.

To calculate the inner products we also need the metric on the surface $ds^2= e^{2 \omega} dz d \bar{z}$ where we are in local coordinates $z$ near the defect. Demanding that the opening angle is $2\pi \a$ at $z_i$ the metric is given by solving Liouville's equation\footnote{We are solving for the constraint $\frac{1}{2}\sqrt{g}\lr{R+2} = 2\pi (1-\a) \delta^2(z-z_i)$ and we have used that $R=-2e^{-2\omega} \hat{\nabla}^2 \omega + e^{-2\omega} \hat{R}$ for a Weyl transformed metric $g = e^{2\omega} \hat{g}$.} with a source
\begin{equation}
    -2 \ol \pd \partial {\omega}+\frac{1}{2}e^{2 \omega}= 2\pi (1-\alpha)\delta^2(z-z_i)\,.
\end{equation}
Solving for the metric perturbatively in $(1-\a)$ we find
\begin{equation}
    \omega=\omega_0 - \frac{1}{2}(1-\alpha)\big[\log|z-z_i|^2+r \big]+\mathcal{O}\lr{\lr{1-\alpha}^2}\,,
\end{equation}
where $\omega_0$ is the Weyl factor for the surface $\Sigma$ without the defect. The linear order term in $(1-\a)$ is split up into a logarithm which gives the delta function singularity, along with a holomorphic function $r$ that satisfies $\ol \pd \partial r = e^{2 \omega_{0} } \log|z-z_i|$.

One final part we need for the calculation of the measure is the beltrami differential $\mu$ that corresponds to infinitesimally moving the defect from $z_i \to z_i + \delta z_i$. As explained in the case of the disk, the beltrami differential can be extracted from a coordinate chart $z'=z+\delta z_i \hspace{.05cm} v(z, \bar z) + \mathcal{O}((\delta z_i)^2)$ such that $z'(z_i) = z_i + \delta z_i$, which implies that $v(z_i, \bar z_i) = 1$. We only need this coordinate chart to linear order in $\delta z_i$. The beltrami differential is then be given by $\mu = \ol \pd v (z,\bar z)$. We will also demand that $v|_{\t{bdy}} = 0$ which is the condition that we are infinitesimally moving the defect but not changing any asymptotically AdS boundaries, which will allow us to integrate by parts. 

We can now calculate all the components of the path integral measure for moving the defect position in $z_i$ coordinates:
\be \label{eqn:muphi_defect}
    \langle \mu,\phi_1\rangle =\int_{\Sigma} d^2 z \hspace{.05cm} \ol \pd v \lr{\frac{1}{z-z_i} + \t{holo.}} = 2\pi,
\ee
where we have integrated by parts and assumed all boundary terms vanish,\footnote{This can be compared to the calculation performed on the disk where we obtain the same answer by performing the full integral \eqref{eqn:diskmuphi}.} used that $\phi_1$ is holomorphic away from the defect, and that $v(z_i,\bar z_i)=1$.\footnote{In our conventions $\ol \pd z^{-1} = 2\pi \delta^2 (z)$ and $\int d^2 z \delta^2 (z) = 1$.}  Computing the inner product of the quadratic differential we find
\be
\begin{aligned}
    \lrb{\phi_1, \phi_1}&=2\int_{\Sigma} d^2 z e^{-2\omega} \left|\frac{1}{z-z_i}+ \ldots \right|^2 = 2 \int_{\Sigma} d^2 z  e^{-2\omega_0} \frac{|z-z_i|^{2(1-\alpha)}}{|z-z_i|^2} + \t{less singular}, \\
    &\underrel{\a \to 1}{=}-\frac{4\pi e^{-2\omega_0(z_i)}}{(1-\alpha)}+\mathcal{O}(1) = -\frac{2\pi}{(1-\a) \sqrt{g(z_i)}} + \mathcal{O}(1).
\end{aligned}
\ee
where in the last line we take $\a \to 1$ and localize onto the most singular part of the integral near $z=z_i$,\footnote{The integral must be performed with a radial cutoff around $z_i$, after which the leading order answer in $(1-\a)$ is cutoff independent.} and in the second equation we notice that we have picked up a factor of the metric $\sqrt{g}$ with the defect removed from the surface, evaluated at $z_i$. Putting everything together, we find that the Weil-Petersson measure \eqref{WP_cncoords} for the integral over the defect position in the blunt angle limit is given by
\begin{equation} \label{eqn:bluntdefect_measure}
   \lim_{\a \to 1} \mathcal{N} \int d^2 z_i \left| \frac{\lb \mu ,\phi_1 \rb}{\sqrt{\lb \phi_1, \phi_1 \rb}} \right|^2 =2 \pi \lr{1-\a} \int d^2 z_i \sqrt{g(z_i)} + \mathcal{O}\lr{(1-\a)^2}.
\end{equation}
Note that on the right $g(z_i)$ is the metric on the surface without the conical defect. From the above we immediately have that the dilaton-gravity operator that inserts a conical defect in the blunt defect limit is given by 
\be
\lim_{\a \to 1} \mathcal{V}_{\a} = 2\pi (1-\alpha) \int d^2 z_i \sqrt{g(z_i)} e^{-2\pi (1-\a) \Phi(z_i)} + \mathcal{O}\lr{(1-\a)^2}. 
\ee
\subsubsection*{Recursion Relation For WP Volumes}
This result also allows us to give a gravitational path integral argument for a recursion relation of Weil-Petersson volumes with blunt defects derived in \cite{Eberhardt:2023rzz}. Consider the Weil-Petersson volume of surfaces $\Sigma$ of genus $g$ with $m$ geodesic boundaries of lengths $\Vec{b}_m = (b_1, \ldots, b_m)$ and $n+1$ conical defects of opening angles $2\pi \alpha_i$ with $\Vec{\alpha}_{n+1} = (\a_1, \ldots, \a_n, \a_{n+1})$. In the limit where one of the defects becomes blunt $\a_{n+1} \to 1$ \cite{Eberhardt:2023rzz} proved the following relation
\be \label{eqn:Turiaci23Recursion}
\frac{d V_{g, m, n+1}\lr{\Vec{\alpha}_{n+1}, \Vec{b}_m}}{d \alpha_{n+1}} \bigg\rvert_{\alpha_{n+1}=1} = -2\pi |\Sigma| V_{g, m, n}\lr{\Vec{\alpha}_n, \Vec{b}_m},
\ee
where $|\Sigma|$ is the hyperbolic area of the surface with $n$ defects satisfying
\be
|{\Sigma}| = -2\pi \lr{2-2g - m -\sum_{i=1}^n (1-\a_i)} = -\frac{1}{2} \int_{\Sigma/\{x_i\}} \sqrt{g} R.
\ee
We can prove this formula be decomposing the volume into an integral over a coordinate parameterizing the position of the $\alpha_{n+1}$ defect, and all the other moduli of the surface. Working near the blunt defect limit the integral over the defect takes the simplified form \eqref{eqn:bluntdefect_measure}
\be
V_{g, m, n+1}\lr{\Vec{\alpha}_{n+1}, \Vec{b}_m} = \int d \lr{\t{other moduli}}
 \times  \lr{2\pi(1-\alpha_{n+1}) \int_{\Sigma/\{z_j\}} d^2 z \sqrt{g} + \mathcal{O}\lr{(1-\a_{n+1})^2} },
\ee
where the integral over the ``other moduli'' take the form of the measure \eqref{WP_cncoords} with various determinants of beltrami and quadratic differentials. The measure for the other moduli implicitly depends on $\alpha_{n+1}$ through the appearance of the defect metric in the inner products defining the measure. However, as we take $\a_{n+1}\to 1$ this dependence goes away since the surface no longer has a defect. Therefore, in the limit that the defect vanishes the integral over the other moduli becomes the Weil-Petersson volume of the surface without the $\a_{n+1}$ defect. Another important point as explained around \eqref{eqn:bluntdefect_measure} is that the metric $\sqrt{g}$ appearing in the above is the metric for the surface without the $\a_{n+1}$ defect. We have also excluded the integral over the points $z_j$ where the other $\vec{\alpha}_n$ defects are located, as these configurations are at the boundary of moduli space.\footnote{This exclusion is necessary to reproduce the relation \eqref{eqn:Turiaci23Recursion} since the Euler characteristic excludes these points.} Taking a derivative we immediately find
\be
\frac{d V_{g, m, n+1}\lr{\Vec{\alpha}_{n+1}, \Vec{b}_m}}{d \alpha_{n+1}} \bigg\rvert_{\alpha_{n+1}=1} = \lr{-2\pi \int_{\Sigma/\{z_j\}} d^2 z \sqrt{g}} \times V_{g, m, n}\lr{\Vec{\alpha}_{n}, \Vec{b}_m},
\ee
which is the desired recursion relation \eqref{eqn:Turiaci23Recursion}. Note that this argument also goes through if we replace the measure for the defect with $\pi \lr{1-\a^2} \int d^2 z \sqrt{g}$, as we found for the disk. 

From the above argument it might be suspected that the volumes also satisfy $\lim_{\a_{n} \to 1} V_{g,n,m} \lr{\Vec{\alpha}_{n}, \Vec{b}_m} \stackrel{?}{=} 0$. In \cite{Eberhardt:2023rzz} this was shown to be true when there are no geodesic boundaries, but is false when such boundaries are present. We do not have a gravity path integral argument for this. One possibility is that with geodesic boundaries there are additional boundary terms that enter into the measure through \eqref{eqn:muphi_defect}, where we assumed that all boundary terms vanished.

\section{Discussion}
In this paper we studied various unresolved aspects of the gravitational path integral of JT gravity. We carried out the gauge fixing of the path integral in second order formalism for general hyperbolic surfaces with asymptotic boundaries and conical singularities. The second order formalism allowed us to clarify the procedure for calculating the proper measure for the conical defect operator, and resolved the question of which dilaton gravity potential should be used for JT gravity coupled to a gas of conical defects \cite{Witten:2020wvy,Maxfield:2020ale,Turiaci:2020fjj,Eberhardt:2023rzz}. This also allowed us to give a gravity path integral argument for certain recursion relations of Weil-Petersson volumes derived using algebraic geometry techniques \cite{Eberhardt:2023rzz}. An open problem is to carry out the full computation of the measure for the conical defect operator to all orders in the $(1-\a)$ expansion on a general surface to prove the conjectured form given in equation \ref{conjecture}, which we were only able to fully compute on the disk topology.\footnote{As explained in the introduction, our reasoning for extending the normalization of the operator found on the disk to arbitrary surfaces is that the operator should be surface independent.}

Along the way we computed determinants of Laplace operators on hyperbolic surfaces with asymptotic boundaries. These determinants are straightforwardly related to partition functions of matter fields minimally coupled to JT gravity. It would be interesting to better understand matter fields coupled to JT with a gas of conical defects, where the bulk geometry would not be pure AdS$_2$. Tangentially, we computed the determinant of the vector Laplacian $\det( \Delta_1+s(s-1))$ on a cone geometry. This can be related to the bulk entanglement entropy of a gauge field on one side of the TFD. Since gauge fields in two dimensions have no propagating degrees of freedom a non-trivial entanglement entropy should arise from edge modes, and it would be interesting to understand this better in AdS$_2$.\footnote{We thank Sean Colin-Ellerin for discussion on this point.} 

We obtained an exact expression for the two-point function of matter fields on an arbitrary surface $\Sigma$ obtainable through the quotient method. The correlator takes into account all geodesics on the surface including self-intersecting geodesics, but does not include an integration over the Schwarizan mode. In \cite{Mertens:2017mtv} diagrammatic rules were derived for correlation functions, including the appearance of $6j$ symbols when geodesics intersect, but this has yet to be derived from the second order perspective. It would be interesting to integrate over the boundary fluctuations and get a closed form expression for the full correlator reproducing the expected $6j$ symbols. This would also allow one to incorporate the contribution of self-intersecting geodesics to the late time two-point function calculation on the handle-disk, where it was argued in \cite{Saad:2019pqd} that such contributions should decay with time.

\subsection*{Acknowledgments}
We thank David Borthwick, Sean Colin-Ellerin, Luca Iliesiu, Geoff Penington, and Joaquin Turiaci for discussion. MU is supported in part by the NSF Graduate Research Fellowship Program under grant DGE1752814, by the National Science Foundation under Grants No. NSF PHY-1748958 and PHY-2309135, by the Berkeley Center for Theoretical Physics, by the DOE under award DE-SC0019380 and under the contract DE-AC02-05CH11231, by NSF grant PHY1820912, by the Heising-Simons Foundation, the Simons Foundation, and National Science Foundation Grant No. NSF PHY-1748958.

\appendix

\section{Consistency of disk measure with general surface} \label{app:diskalphaexpansion}
In this section we explain how the exact calculation for a disk with a single defect is consistent with the measure derived for a general surface. Recall that we found the measure for the defect to be
\begin{equation*}
\mathcal{N}\int d^2 z_i \left|\frac{\lrb{\mu, \phi_1}}{\sqrt{\lrb{\phi_1, \phi_1}}}\right|^2  = \int d^2 z_i ~\times ~\begin{cases}
\pi \lr{1-\a^2} \sqrt{g(z_i)}, &\text{disk,}\\
2\pi \lr{1-\a} \sqrt{g(z_i)} + \mathcal{O}\lr{\lr{1-\a}^2}, &\text{general surface}.
\end{cases}
\end{equation*}
The point is that since the disk calculation is exact we implicitly re-summed the $\lr{1-\a}$ corrections to the measure. We now make this explicit in the case that the defect is at the center. The tower of $\lr{1-\a}$ corrections comes from the inner product for the quadratic differential
\be \label{eqn:appdiskansw}
\lrb{\phi_{1}, \phi_{1}}_{\t{disk}}= \frac{4\pi}{1-\a^2} \frac{1}{\sqrt{g(0)}}, \qquad \lrb{\phi_{1}, \phi_{1}}_{\t{general}} \underrel{\a \to 1}{=} \frac{2\pi}{\lr{1-\a} \sqrt{g(0)}} + \mathcal{O}\lr{1}.
\ee
We can express the metric for the disk with a conical defect in the same form we used for a general surface
\be
ds^2 = e^{2\omega} dz d\bar z, \qquad \omega = \omega_0 + \log \lr{ |z|^{\a-1}}-\log\lr{\frac{1-|z|^{2\a}}{\a^2(1-|z|^2)}}, \qquad \omega_0 = \log\lr{\frac{2}{1-|z|^2}},
\ee
where $\omega_0$ is the Weyl factor for a disk without a defect. We can now calculate the inner product on the disk, doing a series expansion in $(1-\a)$ for the last term in the Weyl factor
\begin{align}
 \lrb{\phi_{1}, \phi_{1}}_{\t{disk}} &=   \int d^2 z \left(\frac{ e^{2\omega_0}}{2}\right)^{-1}  \frac{|z|^{2(1-\a)}}{|z|^2} \times \underbrace{\frac{(1-|z|^2)^2}{\a^2(1-|z|^{2\a})^2}}_{\t{expand}}, \\
 &=\frac{2 \pi}{(1-\a) \sqrt{g(0)}} + \frac{\pi}{\sqrt{g(0)}} + \mathcal{O}\lr{1-\a},\\
 &\stackrel{\tiny \t{re-sum}}{=} \frac{4\pi}{(1-\a^2) \sqrt{g(0)}} .
\end{align}
Which is in agreement with the expression for general surfaces in the $\lr{1-\a}$ expansion, and the re-summation of the series reproduces the full disk answer \eqref{eqn:appdiskansw}.

\section{Double trumpet gluing measure} \label{app:DTmeasure}
The Weil-Petersson measure for gluing two geodesic boundaries is famously given by $\int d b d \tau$ \cite{wolpert1982fenchel,wolpert1983symplectic}, where $b\in [0,\infty)$ is the length of the geodesic boundary and $\tau\in[0,b]$ is a relative twist between the two boundaries being glued. In this appendix we will review how this measure can be derived using Beltrami differentials.

\paragraph{The Beltrami differential.}
We will derive the gluing measure for the double trumpet geometry which can be represented by the quotient of the UHP with metric $ds^2 = \frac{\dz}{\lr{\operatorname{Im}z}^2}$, where the geometry is represented by $\mathbb{H}/\langle T_{b} \rangle$ where $ T_{b}\cdot z = e^{b}z$. The fundamental domain $\mathcal{F}$ is the region between the two semicircles $r=1$ and $r=e^{b}$.
The main idea is to explicitly find the two quasiconformal deformations that infinitesimally deform the geodesic length $b$ and twist $\tau$, and calculate the measure from equation \eqref{WP_cncoords}. The transformation that infinitesimally deforms the length $b \to b + \epsilon$ is given by the map 
\begin{equation}\label{eqn:quasiconfl}
    z\rightarrow f_{b}(z,\bar{z})=z (z \bar{z})^{\frac{\epsilon}{2b}}\,.
\end{equation}
This map satisfies  $ \left|f_{b}({|z|{=}e^{m b}})\right| = e^{m(b+\epsilon)}$, meaning that semicircles with radius $e^{mb}$ are mapped to semicircles with radius $e^{m(b+\epsilon)}$. Thus this transformation increases the geodesic throat size of the double trumpet. For the twist the appropriate deformation is given by
\begin{equation}\label{eqn:quasiconft}
z\rightarrow f_{\tau}(z,\bar{z})=z e^{\epsilon \Phi(\theta)}\,,\qquad \Phi(0)=0,~~ \Phi(\pi)=1\,.
\end{equation}
In the above $\Phi(\theta)$ is an arbitrary smooth function of the angle $z=r e^{i \theta}$. The properties of $\Phi$ ensure that the left boundary of the double trumpet is smoothly infinitesimally twisted relative to the right boundary. For example, a point on the left boundary $z=-1$ is sent to  $z=-(1+\epsilon)$. These maps are known as quasiconformal transformations.\footnote{Quasiconformal transformations satisfy     $|\bar{\partial}f(z,\bar{z})|<\partial f(z,\bar{z})$. A conformal transformation can be viewed as a special class of quasiconformal transformation with $\bar{\partial}f=0$.
Geometrically, a conformal map preserves angles and maps small circles to circles; while a quasiconformal map does not preserve angles and maps small circles to ovals.
Roughly speaking, the eccentricity of the ovals are given by the ratio $\bar{\partial} f/\partial f$. 
The transformation is known as an infinitesimaly quasiconformal transformation if this ratio can be made infinitesimal. Such transformations infinitesimally deform the metric in moduli space, since the new metric is no longer Weyl equivalent to the original metric.}

The Beltrami differential is then defined to be $\mu=\frac{\bar{\partial} f}{\partial f}$.\footnote{Note that the definition of $\mu$ here is the complex conjugate of the definition from most of the math literature.} 
Taking $\epsilon$ to be infinitesimal we can define an infinitesimal beltrami $\hat{\mu}=\lim_{\epsilon\rightarrow 0}\mu/\epsilon$ which through an abuse of notation is the definition of Beltrami differentials used in the main text. 
As explained in the main text, these infinitesimal beltrami differentials capture the infinitesimal change in the metric as we deform the moduli. 
We can directly calculate $\hat{\mu}$ from \eqref{eqn:quasiconfl} and \eqref{eqn:quasiconft} and find 
\begin{equation}
    \hat{\mu}_\tau=\frac{i z}{2\bar{z}}\Phi'(\theta)dz^{-1}d\bar{z}\,, \quad \hat{\mu}_{b}=\frac{z}{2 b \bar{z}} dz^{-1}d\bar{z}\,,
\end{equation}
where we have emphasized that these differentials should be thought of as $(-1,1)$ tensors. It can however be the case that a Beltrami differential does not correspond to a pure deformation of the moduli, for example it can deform the metric by a Weyl transformation. It is thus customary to act on Beltrami differentials by a projection operator that restricts them to pure moduli deformations. This does not affect the calculation of the path integral measure, but ensures that the differentials live in the tangent space to moduli space. The projection is orginally given by \cite{ahlfors1961curvature}
\begin{equation}
P[\hat{\mu}](z)\equiv\frac{6}{\pi}(\operatorname{Im} z)^2 \int_{\mathbb{H}} d^2 \xi\frac{\ol{\hat{\mu}}(\xi,\bar{\xi})}{({\xi}-\bar{z})^4}\,.
\end{equation}
This gives us the projected differentials\footnote{Note that $P[\hat{\mu}_\tau]$ is independent of $\Phi'(\theta)$, which shows that the apparent entire function's worth of degrees of freedom in $\Phi$ do not correspond to genuine moduli deformations.}
\begin{equation}
    P[\hat{\mu}_{b}]=\frac{(\text{Im}z)^2}{{z}^2 b} dz^{-1}d\bar{z}\,, \quad  P[\hat{\mu}_\tau]=\frac{i (\text{Im}z)^2}{\pi {z}^2} dz^{-1}d\bar{z}. 
\end{equation}
One remarkable property of this projection is 
$\langle \hat{\mu},\phi\rangle = \langle P[\hat{\mu}],\phi\rangle$ for any quadratic differential $\phi$, and since we are primarily interested in inner products we use the same notation $\hat{\mu}$ to represent the Beltrami and its projection below. This identity is the statement that the operator projects out the subset of deformations in $\hat{\mu}$ that do not deform the moduli. 

\paragraph{Quadratic differential.}
To compute the metric we also need the quadratic differentials, which are thought of as forming the cotangent space to moduli space. These differentials are quite challenging to compute and were found by Wolpert \cite{wolpert1983symplectic}, we quote the final result\footnote{Note that our convention here is a bit different from Wolpert's convention. The differences are (1) our definition of the $\langle\mu,\phi\rangle$ inner product is twice as Wolpert's and (2) the integral measure on Riemann surface $d^z$ is 2 times the Euclidean measure $ dx dy$ used by Wolpert.}
\begin{equation}\label{eqn:DTphil}
   \phi_b=  \frac{1}{\pi z^2} dz^2 \,, \qquad \phi_\tau=\frac{i}{b z^2}dz^2.
\end{equation}
They have the following inner products with the beltrami differentials $\mu_b, \mu_\tau$\footnote{We slightly change the definition of the inner product by taking the real part, since both $d b$ and $d \tau$ are real variables. } 
\begin{equation}
    \langle \mu_b, \phi_b \rangle \equiv 2\text{Re} \int_{\mathcal{F}} \mu_{b} \bar{\phi}_b=\frac{2}{\pi b}\int_{\mathcal{F}} \frac{(\text{Im}\xi)^2}{\xi^2 \bar{\xi}^2} d^2 \xi =2\,, \qquad \lrb{\mu_\tau, \phi_\tau} = 2,
\end{equation}
where we have used the inner product in \eqref{eqn:muphiprod}. It is straight forward to check that the off-diagonal inner products vanish $\langle \mu_{\tau},\phi_b\rangle = 0$ because of the factor $i$ in the $\mu_{\tau}$ expression.

\paragraph{The Weil-Petersson metric.}
Given these results the Weil-Petersson metric for the moduli space is straightforward to obtain. From the discussion in Section~\ref{sec:3}, we need to evaluate several determinants. 
It is straight forward to calculate 
\begin{equation}
    \det \langle\mu,\phi\rangle=\det \begin{pmatrix}
        \langle \mu_b,\phi_{b} \rangle & \langle \mu_b,\phi_{\tau} \rangle  \\
        \langle \mu_\tau,\phi_{b} \rangle & \langle \mu_\tau,\phi_{\tau} \rangle 
    \end{pmatrix}=\det \begin{pmatrix}
        2 & 0\\
        0 & 2
    \end{pmatrix}=4\,.
\end{equation}
\begin{equation}
  \det \langle\phi,\phi\rangle=\det \begin{pmatrix}
        \langle \phi_b,\phi_{b} \rangle & \langle \phi_b,\phi_{\tau} \rangle  \\
        \langle \phi_\tau,\phi_{b} \rangle & \langle \phi_\tau,\phi_{\tau} \rangle 
    \end{pmatrix}=\det \begin{pmatrix}
        2b & 0\\
        0 & 2 b^{-1}
    \end{pmatrix}=4\,.
\end{equation}

The measure is given by equation \eqref{WP_cncoords}
\begin{equation}
    d\mu_{\text {DT}}= \mathcal{N} \frac{\det \langle \mu,\phi\rangle \det \langle \bar{\mu},\bar{\phi}\rangle}{\det \langle\phi,\phi\rangle} db \, d\tau = 4\mathcal{N} db \, d\tau  \,,
\end{equation}
where the integral ranges over $\tau \in [0,b]$ and $b \in [0,\infty)$. To get the standard form of the gluing measure $\int b d b$ we must choose the normalization $\mathcal{N}=1/4$ for the measure \eqref{ultralocalmeasure} as claimed in the main text.

\section{The determinant calculation} \label{app:determinant}

\subsection*{Notation, conventions, and summary of results}
We first list the notation and main results of the determinant calculation. We denote the surface obtained by a quotient of the Fuchsian group $\Gamma$ as $\Sigma = \mathbb{H}/\Gamma$. The fundamental domain is denoted by $\mathcal{F}_{\Sigma}$ which is thought of as a subregion of $\mathbb{H}$. 
We will sometimes not distinguish between $\mathcal{F}_{\Sigma}$ and $\Sigma$ for simplicity. 
We define the Fuchsian group action to be $T(Z)$ where $T$ is a Fuchsian group element and $Z$ will either be a subregion or a point in $\mathbb{H}$.

An important geometrical invariant of the surface $\Sigma$ is the hyperbolic area
\begin{equation}\label{eqn:hyparea}
    |\Sigma|=-\lr{\frac{1}{2}\int_{\Sigma/\{z_i\}} d^2z \sqrt{g} R + \int_{\partial \Sigma} \sqrt{h} K} =  2 \pi\bigg(2 g+n-2+\sum_{j=1}^k\left(1-n_j^{-1}\right)\bigg),
\end{equation}
where $g$ is the genus, $n$ is the number of boundaries (including cusps and asymptotic boundaries), and $k$ is the number of defects located at positions $z_i$ with deficit angles $\frac{2\pi}{n_j}$. For simplicity we exclude surfaces with geodesic boundaries and cusps ($2\pi$ deficit angles) from the determinant calculation. The area element of $\mathbb{H}$ is $d^2 z \sqrt{g}=y^{-2}dx dy$, and the $u$ variable for hyperbolic distance
\begin{equation}\label{eq:uvar}
    u=\cosh^2\left(\frac{\ell(z,z')}{2}\right)=\frac{1}{2}\left(1+\cosh(\ell(z,z'))\right)=\frac{(x-x')^2+(y+y')^2}{4 y y'}\,,
\end{equation}
where our complex coordinates are given by $z=x+i y$.

We will be interested in the scalar and vector Laplacians. The scalar Laplacian is defined as $-g^{ab}\nabla_a\nabla_b$ acting on scalar functions. Similarly, the vector Laplacian is $\nabla_1^2=-g^{ab}\nabla_a\nabla_b$ acting on two component vectors or covectors, and was introduced in the main text through $\nabla_1^2=\frac{1}{2}P_1^{\dagger}P_1-1$. It will be more convenient to conjugate the vector Laplacian and consider $y^{-2} \nabla_1^2 y^2 + s (s-1)-1=\text{diag}\lr{D_{-1}+s(s-1), D_{1}+s(s-1)}$ where we define the operators 
\be
D_{\pm 1} = -\lr{\partial_x \pm i \partial_y} y^2 \lr{\partial_x \mp i \partial_y},
\ee
where the conjugated operators acts on vectors with top component $v^x+iv^y$ and bottom component $v^x-iv^y$. Conjugated operators have the same spectrum, and so we can equivalently compute the determinant of the above operators. The spectra of $D_{\pm1}$ are identical, and so we find the result that $\det(\Delta_1+1)=\det \lr{D_1+2}^2$. The above immediately implies that $\det(\Delta_1+s (s-1)-1)=\det \lr{D_1+s(s-1)}^2$. We will abuse notation in this appendix and refer to $D_{\pm1}$ as the vector Laplacian.

In the calculation of the determinant we will need the resolvent of the associated operators, also known as the propagator, denoted by $R^0_{\Sigma}$ and $R^1_{\Sigma}$, and their associated traces over the surface ${\mathrm tr} R_{\Sigma} = \int_\Sigma d^2 z \sqrt{g} R_{\Sigma}(z,z)$. Note that the operator, the resolvent, and the trace are defined for a specific surface $\Sigma$. The trace is an integral over the fundamental domain $\mathcal{F}_{\Sigma} \subset \mathbb{H}$.

The main result for the determinants are given as follows \footnote{One can also allow for $n_c$ cusps on the surface, which give a parabolic contribution $Z_{\rm par.}=\left(\Gamma(s-1/2)(2s-1)2^{s-1}\right)^{n_c}$. With the inclusion of cusps, the scalar determinant would include $Z_{\rm par.}$, namely $\det \lr{\ldots} \to \det \lr{\ldots} \times Z_{\rm par.}$ \cite{borthwick2007spectral}. We believe a similar formula holds for the vector determinant.} 
\be\label{eqn:detres0}
    \det\lr{{\Delta}_0+s(s-1)}= {Z_{\text{hyp.}}(s)}{Z_{\text{ell.}}(s)} G_{\infty}(s)^{-\frac{|\Sigma|}{2\pi}} e^{-B s(s-1)+D}\,.
\ee
\be\label{eqn:detres1}
    \det\lr{D_1+s(s-1)}= {Z_{\text{hyp.}}(s)}{Z_{\text{ell.}}(s)} G_{\infty}(s)^{-\frac{|\Sigma|}{2\pi}} (s(s-1))^{\frac{\chi(\Sigma)}{2}}e^{-B s(s-1)+D}\,.
\ee
\be
\det(\Delta_1+s (s-1)-1) = \det\lr{D_1+s(s-1)}^2.
\ee
and the zeta functions are defined by
\begin{equation}\label{eqn:zetas}
\begin{aligned}
    {Z_{\text{hyp.}}(s)} = \prod_{\ell\in \mathcal{L}_{\Sigma}} \prod_{m=0}^{\infty}\left(1-e^{-(s+m)\ell}\right)\,, \quad 
    {Z_{\text{ell.}}(s)} =\prod_{i=1}^{k} \prod_{r=0}^{n_i-1} \Gamma\left(\frac{s+r}{n_i}\right)^{\frac{2 r+1-n_i}{n_i}}\,,
\end{aligned}
\end{equation}
where the explanation for $Z_{\text{hyp.}}$ is given around equation \eqref{eqn:detfinal}.
The function $G_{\infty}$ captures the contribution from the identity element
\begin{equation}
    G_{\infty}(s)=(2 \pi)^{-s} \Gamma(s) G(s)^2\,,
\end{equation}
in which $G(s)$ is the Barnes-G function. 
The constants $B$ and $D$ are given by
\be\label{eqn:constants}
    B=\frac{|\Sigma|}{2\pi},\quad D=\frac{|\Sigma|}{\pi}\zeta'(-1)+\sum_{j=1}^{k}\frac{n_j^2-1}{6 n_j}\log(n_j)\,.
\ee
In the main text we require the calculation of $\det(\Delta_1+1)^{\frac{1}{2}} = \det(D_1+2)$. From the above formulas we immediately find that
\be \label{eqn:DetRatio_Final}
\frac{\det(\Delta_1+1)^{\frac{1}{2}}}{\det(\Delta_0+2)} = 2^{\frac{1}{2}\chi(\Sigma)}.
\ee
\subsection*{The determinant calculation: outline}
The determinant of an operator is divergent since it is the product of infinitely many eigenvalues, and thus a regularization procedure must be specified to calculate determinants. An appropriate regularization procedure is given by specifying the determinant as a solution to the following differential equation\cite{borthwick2007spectral}
\be 
\label{eqn:defdeterminant}
    \frac{1}{2s-1}\frac{d}{ds} \log(\det(\Delta+s(s-1)))= {\mathrm{tr}} R(s) \,,
\ee
where all the operators and resolvents are defined on the surface $\Sigma$. To completely fix the solution we must specify boundary conditions, which are canonically fixed by studying the $s
\to \infty$ behavior of the determinant by the heat kernel method. The calculation of determinants is thus reduced to: (1) the evaluation of the resolvent trace, and (2) understanding the large $s$ asymptotics. 

\paragraph{The resolvent trace}
To calculate the trace of the resolvent we must first obtain the resolvent. For surfaces obtained from the quotient method we can use the method of images to obtain the resolvent. The resolvent for the differential operator $\Delta_0+s(s-1)$ on $\mathbb{H}$ is defined by
\begin{equation}
    (\Delta_0+s(s-1))R_{\mathbb{H}}^{0}(s)=\frac{\delta(x)\delta(y)}{\sqrt{g}}\,.
\end{equation}
The result is given by a function of an $SL(2)$ invariant geodesic length $u(z,w)$
\begin{equation}\label{eqn:r0resolv}
    R^{0}_{\mathbb{H}}(s;z,w)=\frac{1}{4\pi}\frac{\Gamma(s)^2}{\Gamma(2s)} u^{-s}\, {}_2 F_1(s,s,2s;u^{-1})\,.
\end{equation}
The resolvent for the vector Laplacian $D_1+s(s-1)$ is not as trivial. Since the vector Laplacian acts on vectors we require the resolvent to have the following property\cite{teo2021resolvent}
\begin{equation}\label{eqn:r1transf}
    R^1_{\mathbb{H}}\lr{s;T(z),T (w)}T'(z)T'(\ol{w})=R^1_{\mathbb{H}}(s; z, w) \qquad \forall T\in\mathrm{SL}(2)\,.
\end{equation}
We find that 
\begin{equation}\label{eqn:r1resolv}
     R^1_{\mathbb{H}}(s;z,w)=\psi^1(u)\frac{(2i)^2}{(z-\bar{w})^2}\,, \quad  \psi_1(u)=\frac{1}{4\pi}\frac{\Gamma(s-1)\Gamma(s+1)}{\Gamma(2s)}u^{1-s}\,{}_2F_1(s{-}1,s{+}1,2s;u^{-1})\,\,,
\end{equation}
which satisfies the required differential equation
\begin{equation}\label{eqn:d1form}
     (D_1 +s(s-1))R^1_{\mathbb{H}}(s;z,w)=\delta(x)\delta(y)\,.
\end{equation}
Having the resolvent on $\mathbb{H}$, we now proceed to construct the resolvent on $\Sigma$. The key requirement is to enforce the correct transformation property of $R$ under the Fuchsian group action. We get
\begin{equation}
    R_{\Sigma}^0(s;z,w)=\sum_{\gamma\in\Gamma} R^0_{\mathbb{H}}(s;\gamma(z),w)\,,\quad R_{\Sigma}^1(s;z,w)=\sum_{\gamma\in\Gamma} R^1_{\mathbb{H}}(s;\gamma(z),w)\gamma'(z)\,.
\end{equation}
To calculate the resolvent trace we must consider integrals of the diagonal element $R(z,z)$ over the fundamental domain. As the fist steps,  we take $\Gamma$ to be a cyclic group generated by $\gamma_0$. 
Concretely, we define 
\begin{equation}\label{eqn:cyclictr}
\begin{aligned}
    &{\rm tr}R^{0}_{\mathbb{H}/\langle \gamma_0 \rangle }(s)=\sum_{m\neq0}\int_{\mathbb{H}/\langle \gamma_0 \rangle} d^2z \sqrt{g} R^{0}_{\mathbb{H}}(s;\gamma_0^{m}(z),z)\,,\\
    &{\rm tr}R^{1}_{\mathbb{H}/\langle \gamma_0 \rangle }(s)=\sum_{m\neq0}\int_{\mathbb{H}/\langle \gamma_0 \rangle} \frac{d^2z}{2} |\gamma_0'(z)|^{2 m} 
 R^{1}_{\mathbb{H}}(s;\gamma_0^{m}(z),z)\,.
\end{aligned}
\end{equation}
Note that the definitions for the two resolvents are different, due to their different property under coordinate transformation.
The $m=0$ contribution corresponds to the identity, and will need to be handled independently for regularization reasons. 

Evaluating these integrals requires specifying the type of the element $\gamma_0$.\footnote{The parabolic contribution is also easy to calculate, although not used in this paper, with the result ${\rm tr}R_{\mathbb{H}/\langle \gamma_p \rangle }=-\frac{1}{2s-1}\bigg(\log 2+\frac{\Gamma'(s+\frac{1}{2})}{\Gamma(s+\frac{1}{2})}\,\bigg)+\frac{1}{(2s-1)^2}+\text{identity piece}\,.$
}
\begin{enumerate}
[topsep=3pt,itemsep=3ex,partopsep=1ex,parsep=1ex]
    \item hyperbolic $\gamma_h$: the surface $\mathbb{H}/\langle \gamma_h \rangle$ is a double trumpet (with the length of the closed geodesic $\ell$)
    \begin{equation}\label{eqn:hyptraces}
    \begin{aligned}
         & \text{tr}R^0_{\mathbb{H}/\langle \gamma_h \rangle }(s)= \sum_{m\neq0}\int_{\mathbb{H}/\langle \gamma_h \rangle }  {d^2z} \sqrt{g}  R^0_{\mathbb{H}}(s;\gamma_h^m (z),z)=\frac{2\ell}{2s-1}\sum_{m=1}^{\infty} \frac{e^{-sm\ell}}{1-e^{-m\ell}}\,,         \\
         &\text{tr}R^1_{\mathbb{H}/\langle \gamma_h \rangle }(s)= \sum_{m\neq0}\int_{\mathbb{H}/\langle \gamma_h \rangle }  \frac{d^2z}{2}  R^1_{\mathbb{H}}(s;\gamma_h^m (z),z)=\frac{2\ell}{2s-1}\sum_{m=1}^{\infty} \frac{e^{-sm\ell}}{1-e^{-m\ell}}\,,
    \end{aligned}
    \end{equation}
    where the factor of $2$ is from reducing the sum to be over positive integers instead of all non-zero integers.
    \item elliptic $\gamma_e$: the surface is a cone with opening angle $\frac{2\pi}{n}$
    \begin{equation}\label{eqn:elltraces}
    \begin{aligned}
         & \text{tr}R^0_{\mathbb{H}/\langle \gamma_e \rangle }(s)= \sum_{m=1}^{n-1}\int_{\mathbb{H}/\langle \gamma_e \rangle }  {d^2z} \sqrt{g}  R^0_{\mathbb{H}}(s;\gamma_e^k(z),z)=\sum_{m=1}^{n-1} \frac{1}{n\sin \frac{m\pi}{n}} \sum_{t=0}^{\infty}\frac{\sin(2t+1)\frac{m\pi}{n}}{(2s-1)(s+t)}\,,         \\
         &\text{tr}R^1_{\mathbb{H}/\langle \gamma_e \rangle }(s)= \sum_{m=1}^{n-1} \int_{\mathbb{H}/\langle \gamma_e \rangle }  \frac{d^2z}{2}  R^1_{\mathbb{H}}(s;\gamma_e^m(z),z)=\frac{1-n^{-1}}{2s(1-s)} + \sum_{m=1}^{n-1} \frac{1}{n\sin \frac{m\pi}{n}} \sum_{t=0}^{\infty}\frac{\sin(2t+1)\frac{m\pi}{n}}{(2s-1)(s+t)}\,.
    \end{aligned}
    \end{equation}
\end{enumerate}
The identity component must be appropriately regularized by discarding the divergent contribution from the trace, after doing so we find
\begin{align} \label{eqn:identityIntegral}
     \int_{\Sigma}  {d^2z} \sqrt{g}  R^0_{\mathbb{H}}(s;z,z)=\frac{|\Sigma|}{2\pi}\frac{\Gamma'(s)}{\Gamma(s)}+B, \qquad \int_{\Sigma}  \frac{d^2z}{2}  R^1_{\mathbb{H}}(s;z,z)=\frac{|\Sigma|}{2\pi}\left(\frac{\Gamma'(s)}{\Gamma(s)}+\frac{1}{2s(s-1)}\right)+B'
\end{align}
where $B,B'$ will be fixed by boundary conditions. 

Ultimately, the integrals we would like to evaluate are given by
\be \label{eqn:basicResolventIntegral}
\mathrm{tr} R^0(s) = \sum_{\gamma\in\Gamma} \int_{\Sigma} d^2 z \sqrt{g} R^0_{\mathbb{H}}(s;\gamma(z),w), \qquad \mathrm{tr} R^1(s) = \sum_{\gamma\in\Gamma} \int_{\Sigma} \frac{d^2z}{2} |\gamma'(z)|^{2} R^1_{\mathbb{H}}(s;\gamma(z),w).
\ee

Although the sum over the group seems complicated, it simplifies when decomposed into a summation over conjugacy classes. This is because the summation over conjugacy classes can be used to unwrap the integral into the fundamental domain of either the double trumpet or the cone, where it can be easily evaluated. This can be understood as follows. Consider two elements which are conjugate to each other as $\gamma_1=T^{-1} \cdot \gamma_0 \cdot  T$. By using the properties of the resolvent and a change of variables we find 
\begin{equation}
    \int_{\Sigma} d^2z \sqrt{g} R^{0}_{\mathbb{H}}(s;\gamma_1(z),z)= \int_{T(\Sigma)}  d^2z \sqrt{g} R^{0}_{\mathbb{H}}(s;\gamma_0(z),z) \,.
\end{equation} 
Now suppose $\gamma_0$ is a primitive element, and perform a summation over the conjugacy class of $\gamma_0$, denoted by distinct elements $\gamma_i = T_i^{-1} \cdot \gamma_0 \cdot T_i$.\footnote{Here $T_i$ runs over $\Gamma/\langle \gamma_0 \rangle$ so that $\bigcup_i T_i(\Sigma)$ is $\mathbb{H}/\langle\gamma_0\rangle$. } One finds
\begin{equation}
     \sum_{\gamma_{i}}\int_{\Sigma} d^2 z \sqrt{g} R^0_{\mathbb{H}}(s;\gamma_{i}(z),z)= \int_{\bigcup_i T_i(\Sigma)}\sqrt{g} R^0_{\mathbb{H}}(s;\gamma_0(z),z)=\int_{\mathbb{H}/\langle\gamma_0\rangle}\sqrt{g} R^0_{\mathbb{H}}(s;\gamma_0(z),z)\,.
\end{equation}
The same argument works for powers of the primitive element $\gamma_i^{m}$, and so we find that this brings us to the calculation of \eqref{eqn:cyclictr}. Putting everything together we find
\begin{align}
   {\rm tr} R_\Sigma(s)=\underbrace{\sum_{\gamma_h \in \Pi_{h}} \sum_{m \neq 0} \int_{\mathbb{H}/{\langle \gamma_0 \rangle }}  \hskip -18pt R_{\mathbb{H}}\big(s;\gamma_h^{m}(z), z\big)}_{\t{hyperbolic conjugacy classes}}+\underbrace{ \sum_{\gamma_e \in \Pi_{e}} \sum_{m =1}^{n_i-1} \int_{\mathbb{H}/{\langle \gamma_0 \rangle }} \hskip -18pt R_{\mathbb{H}}\big(s;\gamma_e^{m}(z), z\big)}_{\t{elliptic conjugacy classes}}+
   \underbrace{\int_{\Sigma} R_{\mathbb{H}}(s;z,z)}_{\t{identity}} ,
\end{align}
where $R$ is either $R^0$ or $R^1$ with the appropriate measure from \eqref{eqn:basicResolventIntegral}, and $\Pi_{h}$ and $\Pi_{e}$ are representatives of conjugacy classes of generators of maximal hyperbolic and elliptic subgroups of $\Gamma$ respectively. The summation over $k$ for the hyperbolic contribution is over windings of the geodesic, and runs over non-zero integers. For elliptic elements the summation runs from $1,\ldots, n-1$ where $n$ is the order of the elliptic element. If we had cusps we would also include a summation over parabolic subgroups.

 Using the earlier stated integrals, we find
\be
{\rm tr} R^0_\Sigma(s) = \underbrace{\sum_{\ell \in \mathcal{L}_\Sigma} \frac{2\ell}{2s-1}\sum_{m=1}^{\infty} \frac{e^{-sm\ell}}{1-e^{-m\ell}}}_{\t{closed geodesics}} + \underbrace{\sum_{i=1}^k \lr{\sum_{j=1}^{n_i-1} \frac{1}{n_i\sin \frac{k\pi}{n_i}} \sum_{t=0}^{\infty}\frac{\sin(2t+1)\frac{k\pi}{n_i}}{(2s-1)(s+t)}}}_{\t{$k$ conical defects}}+ \underbrace{\frac{|\Sigma|}{2\pi}\frac{\Gamma'(s)}{\Gamma(s)}+B}_{\t{identity}}.
\ee
\begin{align}
 {\rm tr} R^1_\Sigma(s) &= \underbrace{\sum_{\ell \in \mathcal{L}_\Sigma} \frac{2\ell}{2s-1}\sum_{m=1}^{\infty} \frac{e^{-sm\ell}}{1-e^{-m\ell}}}_{\t{closed geodesics}} + \underbrace{\sum_{i=1}^k \lr{\frac{1-n_i^{-1}}{2s(1-s)} + \sum_{j=1}^{n_i-1} \frac{1}{n_i\sin \frac{k\pi}{n_i}} \sum_{t=0}^{\infty}\frac{\sin(2t+1)\frac{k\pi}{n_i}}{(2s-1)(s+t)}}}_{\t{$k$ conical defects}}\nn\\&
 + \underbrace{\frac{|\Sigma|}{2\pi}\left(\frac{\Gamma'(s)}{\Gamma(s)}+\frac{1}{2s(s-1)}\right)+B'}_{\t{identity}}.   
\end{align}
Using ${\mathrm{tr}} R(s)=\frac{1}{2s-1}\frac{d}{ds} \log(\det(\Delta+s(s-1))) $ and carrying out some simple manipulations we find that the ``closed geodesics'' term becomes the Selberg-zeta function, while the conical defect term, combined with the identity, becomes the elliptic contribution and the constant term. Note that in the geodesic sum we restricted to positive integers $m$ which automatically takes into account the closed geodesics with opposite orientation due to the prefactor of two. 


\subsection*{The determinant calculation: details}

We first derive the resolvents \eqref{eqn:r0resolv} and \eqref{eqn:r1resolv} and discuss their properties. We begin with the scalar Laplacian. The resolvent for the scalar Laplacian has to be invariant under SL(2) transformations. Therefore, the resolvent must be a function of hyperbolic distance $u$. Under such an assumption, the differential equation becomes a simple ODE and the solution is given by \eqref{eqn:r0resolv}. 

The vector Laplacian is more complicated. As mentioned above, the resolvent should behaves like a (co)vector under Mobius transformations, as given in  \eqref{eqn:r1transf}. One can then check that \eqref{eqn:d1form} is valid in the sense that both sides transform in the same way. The LHS is already known as \eqref{eqn:r1transf} while the RHS transforms as 
\begin{equation}
    \delta^2(\gamma(z)-\gamma(w))=(\gamma'(z))^{-1}(\gamma'(\bar{w}))^{-1}\delta^2(z-w)\,.
\end{equation}
Importantly, $R^1$ is not invariant under coordinate transformations, so it is not a simple function of hyperbolic distance, and we need to try to write down
\begin{equation}
    R^1(s;z,w)=\psi^1(u)H^1(z,w)\,, \qquad H^1(z,w)=\frac{(2i)^{2}}{(z-\bar{w})^2}\,.
\end{equation}
The function $H^1(z,w)$ captures the non-trivial transformation property since
\begin{equation}
   H^1(\gamma(z),\gamma(w))=(\gamma'(z))^{-1}(\gamma'(\bar{w}))^{-1}H_1(z,w)\,.
\end{equation}
Then we can determine $\psi^{1}(u)$ by  solving the Green's function equation. The solution is
\begin{equation}
    \psi^1(u)=\frac{1}{4\pi}\frac{\Gamma(s-1)\Gamma(s+1)}{\Gamma(2s)}(u)^{1-s}\,{}_2F_1(s{-}1,s{+}1,2s;u^{-1})\,.
\end{equation}
We comment that both resolvents have a logarithmic divergence when $\ell(z,w)\rightarrow 0$ (or $u\rightarrow 1$)
\begin{equation}\label{eq:reslogdiv}
\begin{aligned}
    \lim_{w\rightarrow z } R^0(s;z,w)&=\frac{1}{4\pi} \left(\log(u-1)+2\frac{\Gamma'}{\Gamma}(s)+2\gamma\right)\,,\\
    \lim_{w\rightarrow z } R^1(s;z,w)&=\frac{1}{4\pi} \frac{1}{(\text{Im}z)^2} \left(\log(u-1)+2\frac{\Gamma'}{\Gamma}(s)+2\gamma+\frac{1}{s(s-1)}\right)\,,
\end{aligned}
\end{equation}
which will be useful when calculating the identity piece of the resolvent trace. We now compile some necessary integrals of the resolvent necessary for the calculation.

Note that in \cite{borthwick2007spectral}, similar integrals have been evaluated for the scalar Laplacian. Here we mostly present the details for integrals contributing to the vector Laplacian. The integral techniques are also adopted from \cite{teo2021resolvent}.

\paragraph{Identity element integral.}
To deal with the identity element contribution a regularization prescription must be used since ${\rm tr}R_{\mathbb{H}}(s;z,z)$ is formally divergent. From \eqref{eq:reslogdiv}, we see that the divergent part is independent of $s$, so we can take a $s$ derivative to remove the divergent part.
Therefore, for a hyperbolic surface $X$ we have 
\begin{equation}\label{eqn:identitypieceintegral}
\begin{aligned}\frac{1}{2s-1}\frac{d}{ds}{\rm tr}[R^0_{\mathbb{H}}(s;z,w)]&=\int_{\Sigma}d^2 z \sqrt{g} \lim_{w\rightarrow z}\frac{1}{2s-1}\frac{d}{ds}R^0_{\mathbb{H}}(s,z,w)\\
    &=\int_{\Sigma}\frac{d^2 z}{({\rm Im}z)^2} \frac{1}{2\pi}\frac{1}{2s-1}\frac{d}{ds}\left(\frac{\Gamma'(s)}{\Gamma(s)}\right)\\
    &=\frac{|\Sigma|}{2\pi}\frac{1}{2s-1}\frac{d}{ds}\left(\frac{\Gamma'(s)}{\Gamma(s)}\right)\,.\\
   \frac{1}{2s-1}\frac{d}{ds}{\rm tr}[R^1_{\mathbb{H}}(s;z,w)]&=\int_{\Sigma}d^2 z \lim_{w\rightarrow z}\frac{1}{2s-1}\frac{d}{ds}R^1_{\mathbb{H}}(s,z,w)\\
    &=\int_{\Sigma}\frac{d^2 z}{({\rm Im}z)^2} \frac{1}{2\pi}\frac{1}{2s-1}\frac{d}{ds}\left(\frac{\Gamma'(s)}{\Gamma(s)}+\frac{1}{2s(1-s)}\right)\\
    &=\frac{|\Sigma|}{2\pi}\frac{1}{2s-1}\frac{d}{ds}\left(\frac{\Gamma'(s)}{\Gamma(s)}+\frac{1}{2s(s-1)}\right)\,.
\end{aligned}
\end{equation}
Solving the differential equation will leave a constant $B$ and the constant is the same for both determinants since the divergent behavior of the resolvents is identical. The integral near the asymptotic boundary diverges and is usually regulated using Hadamard finite part regularization, which in the above integral amounts to replacing the infinite volume by the finite hyperbolic volume $|\Sigma|$.

We further comment that \eqref{eqn:identitypieceintegral} gives back to the $G_{\infty}$ function in the determinant because 
\begin{equation}
    \frac{1}{2s-1}\frac{d}{d s} \log G_{\infty}(s)=\frac{\Gamma^{\prime}}{\Gamma}(s)-1\,.
\end{equation}



\paragraph{Hyperbolic element integral.} We evaluate the integral in \eqref{eqn:hyptraces} as
\begin{equation}
    \int_{\mathbb{H}/\langle \gamma_0 \rangle} \frac{d^2z}{2} R^{1}_{\mathbb{H}}(s;\gamma_0^{m}(z),z)= \int_{1}^{e^{\ell}} \frac{dy }{y^2} \int_{-\infty}^{+\infty} dx \frac{(2i)^2}{(z-e^{m\ell} \bar{z})^2} \psi^1(u(z,e^{m\ell}z)) \,,
\end{equation}
where we take the fundamental domain for the double trumpet to be the infinite strip in the UHP. Using a variable substitution $\kappa=\frac{x}{y}\frac{e^{m\ell}-1}{e^{m\ell}+1}$, the integral becomes
\begin{equation}\label{eq:targetint}
    \frac{(2i)^2e^{m\ell}}{(e^{m\ell}-1)(e^{m\ell}+1)}\int_{1}^{e^{\ell}}\frac{dy}{y}  \int_{\mathbb{R}} d\kappa \frac{1}{(\kappa-i)^2}\psi^{1}\left((1 + \kappa^2)
\frac{(1+e^{m\ell})^2}{4 e^{m\ell}}\right)\equiv \frac{\ell}{e^{m\ell}-1} I^1(m\ell)\,.
\end{equation}
The last integral over $\mathbb{R}$ (denoted as $I^1(m\ell)$ including the prefactors) is given by 
\begin{equation}\label{eq:targetint2}
    I^1(e^{m\ell})=\frac{(2i)^2e^{m\ell}}{e^{m\ell}+1}\int_{\mathbb{R}} d\kappa \frac{1}{(\kappa-i)^2}\psi^{1}\left((1 + \kappa^2)
\frac{(1+e^{m\ell})^2}{4 e^{m\ell}}\right)=\frac{e^{-(s-1) m\ell}}{2s-1}\,,
\end{equation}
which can be evaluated using the Mellin transform as will be explained shortly.
With this result, \eqref{eqn:hyptraces} is manifestly true.
 
We now evaluate \eqref{eq:targetint2} as follows. The first step is to observe that $y^{k}$ is an eigenfunction of $D_1$. This means we know how $R^1$ acts on $y^{k}$. The action of $R^{1}$ on $y^{k}$ looks like a Mellin transform of $I^1(m\ell)$. In the end to find the target integral one just need an inverse Mellin transform. 
\begin{enumerate}
[topsep=3pt,itemsep=3ex,partopsep=1ex,parsep=1ex]
    \item From the expression for $D_1$, we see $y^{t}$ can be treated as a ``generalized" eigenfunction\footnote{It is a ``generalized" eigenfunction because it is not normalizable.}, satisfying 
    \begin{equation}
        (D_1+s(1-s))y^t=(t(t+1)+s(1-s))y^t\,.
    \end{equation}
    Furthermore, since the resolvent is the inverse of the Laplacian we get 
    \begin{equation}
        (D_1+s(1-s))\frac{y^t}{t(t+1)+s(1-s)}=(D_1+s(1-s))\int_{\mathbb{H}} R^1(s,z,z')y^{\prime t} d^2 z'\,,
    \end{equation}
    We draw the conclusion that 
    \begin{equation}
        \int_{\mathbb{H}} R^1(s,z,z')y^{\prime t} d^2 w = \frac{y^t}{t(t+1)+s(1-s)}\,,
    \end{equation}
    under the condition that $\text{Re}(s)$ is sufficiently large.\footnote{We need  $\text{Re}(s)>1/2$ to define $R^1$ acting on $y^t$; also we need $\text{Re}(s)>2+\text{Re}(t)$ to make the integral finite.}

    \item Explicitly writing down the integral relation, we see 
    \begin{equation}
        \int_{\mathbb{H}} \frac{(2i)^2}{(x-x'+i (y+ y'))^2}\psi^1\left(\frac{(x-x')^2+(y+y')^2}{4 y y'}\right)y^{\prime t} d^2 z'= \frac{y^t}{t(t+1)+s(1-s)}\,.
    \end{equation}
    Due to the $x$ translation symmetry one can set $x=0$, and further write down the variable substitution $x'=y(1+v)\kappa,~ y'=y v$, so that the integral relation becomes 
    \begin{equation}
        \int_0^{\infty}  \frac{v^{t}}{1+v} dv  \int_{\mathbb{R}} d \kappa \frac{(2i)^2}{(\kappa-i)^2}\psi^{1}\left((1+u^2)\frac{(1+v)^2}{4 v}\right)=\frac{1}{t(t+1)+s(1-s)}\,,
    \end{equation}
    and one should see the similarity to the integral $I^1$ in \eqref{eq:targetint} by  substituting in $v=e^{m\ell}$. 
    Specifically, one gets
    \begin{equation}
       \{\mathcal{M}\,I^1\}=\int_{0}^{\infty} dv\, v^{t-1} I^1(v)=\frac{1}{1-2s} \left(\frac{1}{s+t}+\frac{1}{s-1-t}  \right)\,,
    \end{equation}
    where $\mathcal{M}$ is the Mellin transform.

    \item Using the inverse Mellin transform $\{\mathcal{M}^{-1}\}$, one gets
    \begin{equation}
        I^1(v)=\frac{1}{2\pi i}\int_{c-i\infty}^{c+i\infty} dt \, 
 v^{-t}  \frac{1}{1-2s}\left(\frac{1}{s+t}+\frac{1}{s-1-t}  \right)\,.
    \end{equation}
    One should be careful about the integration contour. Since we will in the end pick $v$ to be $e^{m\ell}>1$, then we want the $t$ contour to close in the right half of the complex $t$-plane. 
    This picks the pole $t=s-1$. 
    In the end, the residue theorem tells us 
    \begin{equation}
        I^1(v)=\text{Res}_{t=s-1}\left[\frac{v^{-t}}{1-2s}\left(\frac{1}{s+t}+\frac{1}{s-1-t} \right)\right]=\frac{v^{-(s-1)}}{1-2s}.
    \end{equation}
    Taking $v=e^{m\ell}$ gives back \eqref{eq:targetint2}.  
    
\end{enumerate}

\paragraph{Elliptic element integral.} We evaluate the integral in \eqref{eqn:elltraces}. We first observe that the fundamental domain $\mathcal{F}_\Sigma=\mathbb{H}/\langle \gamma_e \rangle $ is a cone that cuts out a segment of the hyperbolic disk of angle $2\pi/n$. If $\kappa \in \mathbb{H}/\langle \gamma_e \rangle$ we can change the fundamental domain by acting with $\kappa$
\begin{equation}
   I^{1}(\theta)= \int_{\mathcal{F}_{\Sigma}} R_{\mathbb{H}/\langle \gamma_e \rangle }^{1}(s; z,z) \frac{d^2z}{2}= \int_{\mathcal{F}_\Sigma} R^{1}_{\Sigma}(s;\kappa(z),\kappa(z)) \frac{d\kappa(z)}{2}= \int_{\kappa^{-1}(\mathcal{F}_\Sigma)} R^{1}_{\Sigma}(s;z,z) \frac{d^2z}{2}\,,
\end{equation}
without changing the integral since we simply rotate the fundamental domain. The trick is to combine $n$ fundamental domains together to get back $\mathbb{H}$, and divide by $n$ to not overcount, such that the integral $I^{1}(\theta)$ becomes 
\begin{equation}
    I^1(\theta)=\int_{\mathbb{H}} \frac{d^2 z}{n}\left(\frac{2i}{\sin\theta(1+x^2+y^2)-2i\cos\theta y}\right)^2\psi^1\left(\frac{\sin ^2 \theta\left(1+x^2+y^2\right)^2+4 \cos ^2 \theta y^2}{4 y^2}\right)\,.
\end{equation}
Since only $x^2$ is involved, we assume $x>0$ and add a factor 2 to the integral.
By using the variable substitution $t=\frac{1+x^2+y^2}{2y}$ (so that $x=\sqrt{2yt -1 -y^2}$, we get 
\begin{equation}
\begin{aligned}
    I^1(\theta)&=-\frac{2}{n}\int_{0}^{\infty} dy  \int_{\frac{1+y^2}{2y}}^{\infty} dt\, \frac{\psi^1(t^2\sin^2\theta+\cos^2\theta)}{(t\sin\theta-i\cos\theta)^2}\frac{1}{y^2} \frac{dx}{dt}\\
    &=-\frac{2}{n}\int_{0}^{\infty} dy  \int_{\frac{1+y^2}{2y}}^{\infty} dt\, \frac{\psi^1(t^2\sin^2\theta+\cos^2\theta)}{(t\sin\theta-i\cos\theta)^2}\frac{1}{ y \sqrt{2yt -1 -y^2}}\,.
\end{aligned}
\end{equation}
Then we switch the order of integral 
\begin{equation}
\begin{aligned}
    I^1(\theta)&=-\frac{2}{n}\int_{1}^{\infty} dt  \int_{t-\sqrt{t^2-1}}^{t+\sqrt{t^2-1}} dy \, \frac{\psi^1(t^2\sin^2\theta+\cos^2\theta)}{(t\sin\theta-i\cos\theta)^2}\\
    &=-\frac{2\pi}{n}\int_{1}^{\infty} dt \, \frac{\psi^1(t^2\sin^2\theta+\cos^2\theta)}{(t\sin\theta-i\cos\theta)^2}\\
    &=\frac{4\pi i}{n}\int_{1}^{\infty} dt \,(t\sin\theta\cos\theta) \frac{\psi^1(t^2\sin^2\theta+\cos^2\theta)}{(t^2\sin^2\theta+\cos^2\theta)^2}  \\
    &\qquad - \frac{2\pi}{n}\int_{1}^{\infty} dt \,(t^2\sin^2\theta-\cos^2\theta) \frac{\psi^1(t^2\sin^2\theta+\cos^2\theta)}{(t^2\sin^2\theta+\cos^2\theta)^2}
\end{aligned}
\end{equation}
The imaginary part is easy to evaluate (here $p=t^2\sin^2\theta+\cos^2\theta$) 
\begin{equation}
    \frac{2\pi i}{n}\cot\theta\int_{1}^{\infty} dp \, p^{-2}  \psi^{1}(p) =\frac{i\cot\theta}{2 n s(s-1)}\,,
\end{equation}
while the real part is quite involved, with the result
\begin{equation}
     \frac{1}{2 n \sin\theta}\left(\frac{\sin\theta}{s(1-s)} + \sum_{t=0}^{\infty}\frac{2\sin(2t+1)\theta}{(2s-1)(s+t)}\right)\,.
\end{equation}
The final answer is 
\be
I^1(\theta) = \frac{i\cot\theta}{2 n s(s-1)} + \frac{1}{2 n \sin\theta}\left(\frac{\sin\theta}{s(1-s)} + \sum_{t=0}^{\infty}\frac{2\sin(2t+1)\theta}{(2s-1)(s+t)}\right).
\ee
To obtain \eqref{eqn:elltraces} we perform the sum over the group $\sum_{m=1}^{n-1} I^1(\frac{2\pi m}{n})$, which kills off the imaginary contribution, and gives the $(1-n^{-1})$ factor in \eqref{eqn:elltraces}.

\paragraph{Determining the constants in the determinants}
The determinant is given in terms of a differential equation, and so we must fix boundary conditions to fix the constants. The constants are fixed by matching to the large $s$ behavior of the determinant computed using the heat kernel method. We introduce 
\begin{equation}\label{eq:heatkernel}
\begin{aligned}
    &\Theta_{i}(t)={\rm tr}\big(e^{-t\Delta_i}\big)\,,\\
    &\log \det(\Delta_i+s(1-s))=-\frac{\partial}{\partial w} \frac{1}{\Gamma(w)} \int_{0}^{\infty}dt t^{w-1}e^{t s(1-s)} \Theta_{i}(t) \Big|_{w=0}\,,
\end{aligned}
\end{equation}
where $\Delta_i$ is any Laplace operator. Since we are taking $s\to\infty$, the integral is dominated by $t=0$, and we can expand the heat kernel ${\rm tr}\big(e^{-t\Delta}\big)$ in this limit where it has universal behavior \cite{teo2021resolvent}
\begin{equation}\label{eq:asympheatkernel}
    \lim_{t \to 0} \Theta(t)\sim \frac{a}{t}+\frac{b\log t}{\sqrt{t}}+\frac{c}{\sqrt{t}}+d+\mathcal{O}(t^{\#}).
\end{equation}
The constants $a,b,c,d$ depend on $\Sigma$ and the operator $\Delta_i$. We find the answer\footnote{The two $t^{-1/2}$ terms come from the parabolic elements and we discard them since we do not consider cusps.}
\begin{equation}
  \lim_{s\to\infty}\log \det(\Delta_i+s(1-s)) = \frac{\partial}{\partial w} \frac{1}{\Gamma(w)} \int_0^{\infty} dt t^{w-1}\mathrm{e}^{-t u^2}\lr{\frac{a}{t}+d} \Big|_{w=0}=-a u^2+ a u^2 \log u^2 - d \log u^2 \,,
\end{equation}
where $u=s-1/2$. The asymptotic expansions as $s\to\infty$ of the different contributions to $\log\det (\Delta_i+s(1-s))$ are
\begin{equation}
\begin{aligned}
    \log G_{\infty}(s)^{-\frac{|\Sigma|}{2\pi}}&=-\frac{|\Sigma|}{2\pi}\left\{\frac{1}{2}u^2 \log u^2-\frac{3}{2} u^2 + \frac{1}{12}\log u+2\zeta^{\prime}(-1)\right\}+\mathcal{O}(1)\\
    \log Z_{\mathrm{ell.}}(s)&=\sum_{j=1}^\nu\frac{m_j^2-1}{6 m_j} ( \log u- \log{m_j})+\mathcal{O}(1)\\
    \log Z_{\rm hyp.}(s)&=\mathcal{O}\lr{\frac{1}{s^{\#}}} \\
    \log e^{B s(1-s)+D}&=-B u^2 +D+ \frac{1}{4}
    \,.
\end{aligned}
\end{equation}
Demanding that terms proportional to $u^2$ and $u^2\log u^2$ have opposite coefficients, and no extra constant term in the expansion, we can fix the  constants $B$ and $D$ to be
\be
    B=\frac{|\Sigma|}{2\pi},\quad D=\frac{|\Sigma|}{\pi}\zeta'(-1)+\sum_{j=1}^{\nu}\frac{m_j^2-1}{6 m_j}\log(m_j)\,,
\ee
as stated earlier, and it turns out the constants are identical for both the scalar and vector Laplacians. 


\bibliographystyle{utphys}
\bibliography{references}
\end{document}